\documentclass[preprint, 1p, authoryear]{elsarticle}

\let\today\relax
\makeatletter
\def\ps@pprintTitle{%
	\let\@oddhead\@empty
	\let\@evenhead\@empty
	\def\@oddfoot{\footnotesize\itshape
		{} \hfill\today}%
	\let\@evenfoot\@oddfoot
}
\makeatother

\usepackage[titletoc,toc,title]{appendix}

\usepackage{stackengine}

\usepackage[capposition=top]{floatrow}

\usepackage{epstopdf} 
\usepackage{bbm, bm}
\usepackage[inline]{enumitem}

\usepackage{makecell}

\usepackage{hvfloat}

\usepackage[hyphens]{xurl}
\usepackage[hidelinks]{hyperref}
\hypersetup{breaklinks=true}
\usepackage{cite}

\usepackage{amssymb,amsmath,amsthm,mathtools}
\usepackage{graphicx}
\usepackage{dsfont}

\usepackage{bigstrut,booktabs,varioref,float,tabularx,multirow}
\usepackage{tablefootnote}

\usepackage{pdflscape}

\usepackage{color}
\usepackage{soul}
\usepackage{adjustbox}
\usepackage{changepage}

\usepackage{afterpage}

\usepackage{tikz}
\usetikzlibrary{positioning}
\usetikzlibrary{arrows}
\usetikzlibrary{shapes.multipart}

\theoremstyle{plain}

\theoremstyle{definition}

\theoremstyle{remark}

\newcommand{\tabref}[1]{Table~\ref{#1}}

\newcommand{\figref}[1]{Figure~\ref{#1}}

\newcounter{TKcommentCounter}

\usepackage{cancel}

\title{{The effect of the COVID-19 health disruptions on breast cancer mortality for older women: A semi-Markov modelling approach}}

\author[1] { Ay\c{s}e Ar{\i}k \corref{cor1}%
}
\ead{A.ARIK@hw.ac.uk}

\author[1] { Andrew J.G. Cairns }

\author[2] {Ereng{u}l Dodd}

\author[1] { Angus S. Macdonald  }

\author[1] { George Streftaris }

\cortext[cor1]{Corresponding author}

\address[1] {Department of Actuarial Mathematics and Statistics, 
Heriot-Watt University, and the Maxwell Institute for Mathematical Sciences, 
UK}
\address[2] {  Mathematical Sciences, S3RI, University of Southampton, 
Southampton, United Kingdom}

\begin{document}

\begin{abstract}

Public health measures necessitated by the COVID-19 pandemic have 
affected cancer pathways by halting screening, delaying diagnostic tests and reducing the
numbers starting treatment. Specifically, this moves individuals from observed and treated 
pathways to unobserved and untreated pathways. We introduce a semi-Markov model 
that includes both, extending an industry-based multiple state model used for life and 
critical illness insurance. Our model includes events related to cancer diagnosis and 
progression based on publicly available population data for women aged 65--89 in England 
and on relevant medical literature. We quantify age-specific excess deaths, for a period up 
to 5 years, along with years of life expectancy lost and changes in cancer mortality by 
cancer stage. Our analysis suggests a 3--6\% increase in breast cancer deaths, and a 4--6\% 
increase in registrations of advanced breast cancer, robust under sensitivity analysis. 
This should be applicable to actuarial work in areas where longevity and advanced age 
morbidity affect healthcare, retirement and insurance.

\end{abstract}

\maketitle


\section{Introduction}

The COVID-19 pandemic has claimed more than 6.2 million lives worldwide as of May 2022 \citep{WHO2021}. 
{\color{black} The pandemic has alerted actuaries, epidemiologists and longevity specialists not only because of the increased number of deaths, but also because of the potential future impact of healthcare disruptions resulting from imposed public health measures.}
During the pandemic the UK entered three national lockdowns, with the first being introduced on 23 March, 2020.
Cancer pathways have been seriously affected by the changes in health practices due to a halt in cancer screening (from late March 2020 till June 2020), {significant increases in the number of patients waiting for key diagnostic tests for more than 6 weeks}, and significant reductions in the number of patients starting cancer treatment.
Cancer Research UK (CRUK) has reported that 3 million fewer people were screened for cancer in the UK between March and September 2020. 
Moreover, the number of cancer patients starting a cancer treatment decreased by 12\% between April 2020 and March 2021 compared to the pre-pandemic levels, 
whereas the number of people waiting for more than 6 weeks for key diagnostic tests has soared to 215,000 in March 2021 from 67,000 in March 2020 \citep{CRUKCIT2021}. 
These figures sparked fear of a shift to later diagnosis for people having the disease but not diagnosed yet. 
This could restrict the opportunities for feasible treatment and worsen cancer survival. 
{\color{black} This has also triggered concerns regarding changes in cause-specific mortality, e.g. from cancer, impacting trends of all-cause mortality.}

{\color{black} To the best of our knowledge, this is the first actuarial study developed to quantify the impact of healthcare disruptions on cancer mortality.}
\citet{Laietal2020} point out dramatic reductions in the demand for, and supply of, cancer services in response to the COVID-19 pandemic by showing that these reductions could increase excess mortality among cancer patients.
\citet{Sudetal2020} indicate a significant reduction in cancer survival as a result of treatment delay, mostly disruption in cancer surgery.
\citet{Maringeetal2020} also note substantial increases in avoidable cancer deaths 
in England as a result of diagnostic delays of over a year. 
\citet{Ariketal2021} report significant increases in type-specific cancer mortality as a result of diagnostic delays.
\citet{Alagozetal2021} project a small long-term cumulative impact on breast cancer (BC) mortality in the US over the next decade due to initial pandemic-related disruptions.

Early empirical studies suggested that COVID-19 is more likely to affect older people and those with comorbidity \citep{Chenetal2020, Richardsonetal2020, Grassellietal2020, Zhouetal2020}. 
{\color{black}Furthermore, developing COVID-19 has been shown to be a greater risk for cancer patients 
depending on type of malignancy, age, and gender \citep{Pinatoetal2020, Garassinotal2020, Leeetal2020, Sainietal2020}. 
\citet{Pinatoetal2021} reported that cancer patients in the UK have been more severely affected by the COVID-19 pandemic compared to those in continental Europe. }

Part of the contribution of this work is providing a modelling framework, which goes beyond the aforementioned empirical work, to investigate the impact of a pandemic, such as COVID-19, on BC mortality and morbidity.
{\color{black} Our modelling approach also allows the extension of a critical illness model widely applied by the insurance industry \citep{ReynoldsFaye2016}, to include states relevant to cancer progression and undiagnosed cases, and to 
 account for changes in diagnostic and treatment services . 
Thus, the proposed modelling framework can also be implemented by the insurance industry in the context of critical illness and life insurance applications as demonstrated by \citet{Ariketal2023b}.}
Particularly, we are interested in how the pandemic, causing major disruption to the health service, may affect mortality associated with disorders normally treated by the health service.
It is assumed that the pandemic may give rise to changes by preventing or delaying the detection or diagnosis of BC.
We examine the impact of diagnostic delays up to 5 years, as \citet{Maringeetal2020} state that `the effect of delayed presentation on patients with cancer is not immediate, and premature death as a result might occur up to 5 years later \ldots' (p. 1024). 
This is motivated by screening programmes and cancer treatments having been largely affected by lockdowns.
{\color{black} This is relevant and important to insurers since BC remains one of the most common conditions amongst female critical illness claims \citep{CMI2011, Aviva2015}. 
It is worth to noting that a screening programme is available for BC, which plays a crucial role for early diagnosis and increases the chances of BC survival \citep{CRUK2020}. However,} according to \citet{CRUKCIT2021}, 
7,200 fewer cases of BC were diagnosed between April--December 2020 compared to the same period in 2019, 60\% fewer cases were diagnosed via screening, 
whilst 22\% fewer patients started treatment from April 2020 till March 2021, compared with the same period in 2019.

Quantifying the impact of cancer diagnosis delays by considering cancer stage is complex in the light of insufficient data, but a Markov approach provides 
a suitable modelling framework \citep{Castellietal2007, Luetal2011, Adamsetal2013, Buchardtetal2015, Hubbardetal2016, BaioneandLevantesi2018, Hacarizetal2021, Soeteweyetal2022}. 
We establish a {semi-Markov} model with multiple states, including observed and unobserved BC cases, based on: \begin{enumerate*}[label=(\roman*)]
	\item available cancer registration and deaths data in England, provided by  the Office for National Statistics (ONS); and
	\item published clinical studies.
\end{enumerate*}
{\color{black} Our approach presents a more detailed model of BC as compared to the multiple state model introduced in a study 
by the UK Continuous Mortality Investigation Committee (\citet{CMI1991}), and the critical illness insurance model in \citet{ReynoldsFaye2016}, adopted by the insurance industry.
Our model differs from the earlier literature models in two important ways: 
\begin{enumerate*}[label=(\roman*)]
\item by differentiating between observed and unobserved cancer cases; and 
\item by introducing cancer stage information.
\end{enumerate*} }
Accordingly, we estimate age-specific, short-term excess deaths, in addition to years of life expectancy lost (YLL) from cancer, with particular emphasis on ages above 65. 

This paper is organised as follows. 
In Section~\ref{sec:Methodology} we introduce the model for BC risk.
In Section~\ref{seca:Results} we calibrate the model in a pre-pandemic {environment}. 
In Section~\ref{secb:Results} we introduce two `pandemic' scenarios. 
In Section~\ref{secc:Results} we estimate excess deaths and YLLs under a pre-pandemic model calibration and pandemic scenarios. 
In Section~\ref{Sec:SensitivityAnalysis} we provide a sensitivity analysis for model assumptions. 
Finally, in Section~\ref{sec:Conclusion} we discuss our findings and their implications along with strengths and limitations of our approach.

\section{Methodology} \label{sec:Methodology}

\subsection{Definitions of breast cancer stages} \label{sec:MarkovModel}

{BC mortality is the most common cancer diagnosed in women, 
in addition to being one of the leading causes of death for women \citep{ONSPHE2019, PHE2017CoD}. 
	The most common type of BC is known to be `invasive' BC that indicates cancer cells spreading from the ducts into the surrounding (breast) tissues, with the two most well-known ones are `invasive ductal carcinoma' and `invasive lobular carcinoma'. Invasive BC can be described from early to advanced stage BC \citep{CRUKInvasiveBC2020}. }The clinical model of BC progression is a well-defined staging model of the form:

\begin{eqnarray*}
	\mbox{No BC} \rightarrow \mbox{Stage 1 BC} \rightarrow \mbox{Stage 2 BC} \rightarrow \mbox{Stage 3 BC} \rightarrow \mbox{Stage 4 BC} \rightarrow \mbox{Dead from BC} 
	\label{eq:BCstages}
\end{eqnarray*}

\noindent {where a higher stage number shows that cancer tumour is bigger or has spread from breast to distant parts of the body, also known as `metastasis'. This staging model, namely TNM, categorises cancer from Stage 1 to Stage 4 based on the tumour (T) size, that can be between 1--4 with 1 for small tumours and 4 for large tumours; whether or not lymph nodes (N) have cancer cells, that can change between 0--3; and whether or not the cancer cells move to other parts of the body (M), that can be either 0 or 1 \citep{ONS2017CancerSurvival, CRUKCancerStage2020}. 
}

The progression from Stage 1 to Stage 4 is assumed to be real and physical, whether observed or not. It is possible that `transition into Stage 1 BC' is the nearest equivalent in the model to `onset of BC'. We assume that `dead from BC' is accessible only from Stage 4, and `dead from other causes' (not shown above) is accessible from all `live' states. 

\medskip
The clinical staging model {above} takes no account of what is observed or unobserved, i.e. all women free of BC and dead from BC are observed. {In reality, }an individual in one of BC Stages 1--4 may be observed to be so, or unobserved, represented by separate states. Transitions are possible:

\begin{itemize}
	\item forward through stages of BC; and
	\item from `No BC' or an unobserved BC state to an observed BC state.
\end{itemize}

The latter possibility we take to be the same as `diagnosis' {event, that is the first occurrence of BC observed}. Thus a woman who is diagnosed with Stage 3 BC makes a transition from either `Stage 2, Unobserved' or `Stage 3, Unobserved' to `Stage 3, Observed' and so on.


\subsection {Modelling unobserved breast cancer}\label{Sec:BCModel}

We distinguish BC death from other causes of death and define life histories accordingly, keeping in mind that the main focus of this work is {providing a methodology }on quantifying the impact of BC diagnostic delays.

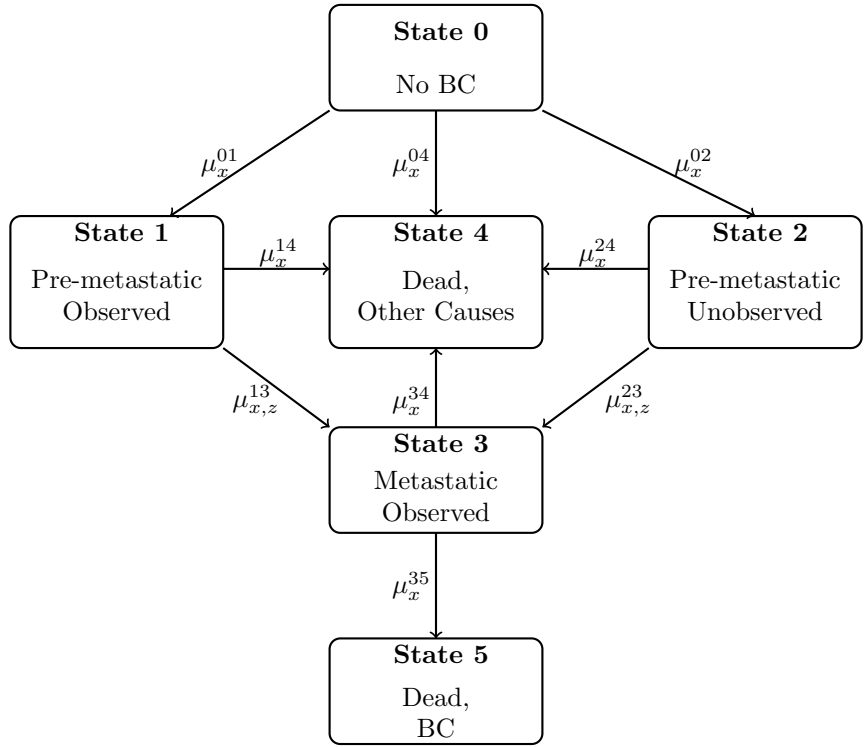
\begin{figure}[H]
	\begin{center}
		\begin{tikzpicture}[scale=.7]
			\draw[rounded corners, thick] (6,4) rectangle (10,6); 
			\node at (8,5.5) {\textbf{ State 0}};
			\node at (8,4.5) {No BC};
			
			\draw[rounded corners, thick] (0,-0.5) rectangle (4,2); 
			\node at (2,1.7) {\textbf{ State 1}};
			\node[align=center] at (2,0.5) {Pre-metastatic\\ Observed};
			
			\draw[rounded corners, thick] (12,-0.5) rectangle (16,2); 
			\node at (14,1.7) {\textbf{ State 2}};
			\node[align=center] at (14,0.5) {Pre-metastatic\\ Unobserved};
			
			\draw[rounded corners, thick] (6,-0.5) rectangle (10,2); 
			\node at (8,1.7) {\textbf{ State 4}};
			\node[align=center] at (8,0.5) {Dead,\\ Other Causes};
			
			\draw[rounded corners, thick] (6,-4) rectangle (10,-2); 
			\node at (8,-2.3) {\textbf{ State 3}};
			\node[align=center] at (8,-3.3) {Metastatic\\ Observed};

			\draw[rounded corners, thick] (6,-8) rectangle (10,-6); 
			\node at (8,-6.3) {\textbf{ State 5}};
			\node[align=center] at (8,-7.4) {Dead,\\ BC};
			
			\draw[->, thick] (6,4) -- (3,2); 
			\node[right] at (3.4,3) {$\mu^{01}_{x}$};
			\draw[->, thick] (10,4) -- (14,2); 
			\node[right] at (12.3,3) {$\mu^{02}_{x}$};
			\draw[->, thick] (8,4) -- (8,2); 
			\node[right] at (7,3) {$\mu^{04}_{x}$};
			\draw[->, thick] (4,1) -- (6,1); 
			\node[right] at (4.5,1.3) {$\mu^{14}_{x}$};	
			\draw[->, thick] (12,1) -- (10,1); 
			\node[right] at (10.5,1.3) {$\mu^{24}_{x}$};	
			\draw[->, thick] (8,-2) -- (8,-0.5); 
			\node[right] at (7,-1.5) {$\mu^{34}_{x}$};	
			\draw[->, thick] (8,-4) -- (8,-6); 
			\node[right] at (7,-5) {$\mu^{35}_{x}$};	
			\draw[->, thick] (4,-0.5) -- (6,-2); 
			\node[right] at (4,-1.5) {$\mu^{13}_{x,z}$};	
			\draw[->, thick] (12,-0.5) -- (10,-2); 
			\node[right] at (11,-1.5) {$\mu^{23}_{x,z}$};	
			
		\end{tikzpicture}
	\end{center}
	\caption{A breast cancer semi-Markov model in continuous time. 
		\label{fig:BC_model}}
		\floatfoot{Note: Intensities $\mu$ may be functions of age $x$ and/or duration $z$.
		}
\end{figure}

{In \figref{fig:BC_model}, we introduce a model of BC progression, based on the stages described in Section~\ref{sec:MarkovModel}, 
but introducing some simplifications (Section~\ref{Sec:BCModelAssumptions}) based on the available data and published clinical studies (Section~\ref{seca:Results}). }
\figref{fig:BC_model} shows a schematic representation of a continuous-time model for the life history of a woman at age $x$.
Age-specific transition intensities from state $i$ to state $j$ are denoted by $\mu_{x}^{ij}$, where $x$ is age-at-entry to state $i$.
{Age- and duration-dependent transition intensities at age $x$ and duration $z$ from state $i$ to state $j$ are denoted by $\mu^{ij}_{x, z}$.}
Stages 1, 2 and 3 of BC combined are represented by States 1 and 2 in the model, State 1 being observed cases and State 2 being unobserved cases. All stage 4 cases of BC are represented by State 3 of the model, and are assumed 
to be observed. 
{We note here that `stage' and `state' are distinct concepts in this paper.}

{In the semi-Markov model considered here, \figref{fig:BC_model}, the usual Kolmogorov equations in a Markov model
	are replaced by a system of integral-differential equations, with integrals over duration being required for certain states. 
	Often such integrals can be intractable. In our model, which has no more than one duration-dependent transition in any possible life history, the required integrals are of low dimension and the modified Kolmogorov equations can be solved numerically using standard methods (Appendix~\ref{sec:AppendixKolmogorov}). 
	In particular, we apply a fourth-order Runge-Kutta scheme to solve the modified Kolmogorov equations under consideration \citep{Macdonaldetal2018}.
}

\subsection {Modelling assumptions}\label{Sec:BCModelAssumptions}

We introduce the following modelling assumptions.

\begin{itemize}
	\item[{\bf A1}:] States 1 and 2 both represent Stages 1--3 of BC progression. We do not attempt to model progression between these stages explicitly as this is not supported by available data. 
	Note that Stages 1--3 BC have a similar pattern for one-year survival \citep{ONS2016CancerSurvival}. 
	State 3 represents Stage 4 of BC progression. 
	This accords with assumptions in some epidemiological studies \citep{Zhaoetal2020}. 
	\item[{\bf A2}:] State 5 (`Dead, BC') is accessible only from State 3 (`Metastatic BC'). That is, earlier stages of BC lead to death from BC only by first progressing to metastasis.
	\item[{\bf A3}:] {All individuals entering State 3 are observed to do so}, whether their progression prior to entering  that state was observed or not. That is, death from BC without metastatic BC being noticed pre-mortem is rare enough to ignore \citep{RedigandMcAllister2014}.	
\end{itemize}

The model also includes a state representing unobserved cases of BC, State 2 (`Pre-metastatic Not Observed').
With the pandemic shock in mind, for the purpose of modelling \emph{changes} in BC mortality caused by {dramatic} changes in the health service, we add two more model assumptions relating to State 2:

\begin{itemize}
	\item[{\bf A4}:] Neither the manner in which we observe BC, nor the presence of a pandemic, affect the overall new cases of cancer. Therefore, we assume the total transition from `No BC' to BC stays constant. 
	That is  
	\begin{equation}
	\mu_x^{01} + \mu_x^{02} = \mu_x^*,
		\label{eq:BCDiagUnchnaged}
	\end{equation}

	where $ \mu_x^*$ is independent of any particular pandemic scenario.
	\item[{\bf A5}:] Individuals in State 1 (`Pre-metastatic Observed') are assumed to be treated for BC, while individuals in State 2 are assumed not to be treated. Therefore, we assume $\mu_{x,z}^{13} < \mu_{x,z}^{23}$ for the same age. Moreover, we assume that treatment given while in State 1, e.g. the type of treatment, does not depend on any particular pandemic scenario, so the transition intensities $\mu_{x,z}^{13}$ and $\mu_{x,z}^{23}$ also do not depend on any particular pandemic scenario.
\end{itemize}

A4 and A5 suggest a convenient parametrisation of the model:

\begin{equation}
	\mu_x^{01} = \alpha_x \, \mu_x^*, \qquad \mu_x^{02} = (1 - \alpha_x) \, \mu_x^*,   \qquad  \mu_{x,z}^{13} = \beta_{x,z} \, \mu_{x, z}^{23} \qquad (\beta_{x,z} < 1),
	\label{eq:Parametrisation}
\end{equation}

\noindent where $0<\alpha_x<1$ quantifies the proportional relationship between $\mu^{01}_x$ and $\mu^{02}_x$, and will later be used to determine pandemic scenarios. 
For simplicity, and lacking data to support other assumptions, 
we assume $\alpha_x = \alpha$ and $\beta_{x,z}=\beta$. 
We suppose that $\mu_{x,z}^{23}$ represents the rate of progression to metastatic BC in the absence of treatment, and $\beta$ measures the effectiveness of treatment. So, our approach assumes that $\mu_x^*$ and $\beta$ are fixed regardless of any pandemic scenario.

\section{Calibration of the Model} \label{seca:Results}

The model is calibrated based on the population of women in England, in age groups 65--69, $\ldots$, 85--89. 
These population estimates are the closest we have to represent the exposure in State 0 that are women to be free of BC. 
{However, we note that these estimates do not distinguish whether or not a woman is actually free of BC, leading to a potentially higher exposure in State 0.}
The aim is to estimate occupancy probabilities for each model state at future times. Calibrating the model means estimating the distribution by age in State 0 between 1 January 2020 and 31 December 2024, and the transition intensities in the model. We rely on published clinical studies and a set of cancer data collected by the ONS. 
We describe the sources we use in the following sections.

{\subsection{Available data: Population incidence and mortality rates of breast cancer} \label{sec:ONSData}}

We consider new cancer diagnoses/registrations and deaths data between 2001--2017 in England, provided by the ONS. Cancer registrations are split by five-year age groups (20--24, 25--29, \ldots, 85--89), type of tumour, single calendar year, and gender. 
{Causes of death data have similar granularity, up to 2018.}
Corresponding mid-year population estimates are available from the ONS. 

\begin{figure}[H]
	\subfloat[Breast cancer incidence \label{fig:BreastIncidence}]{\includegraphics[width=0.5\textwidth]{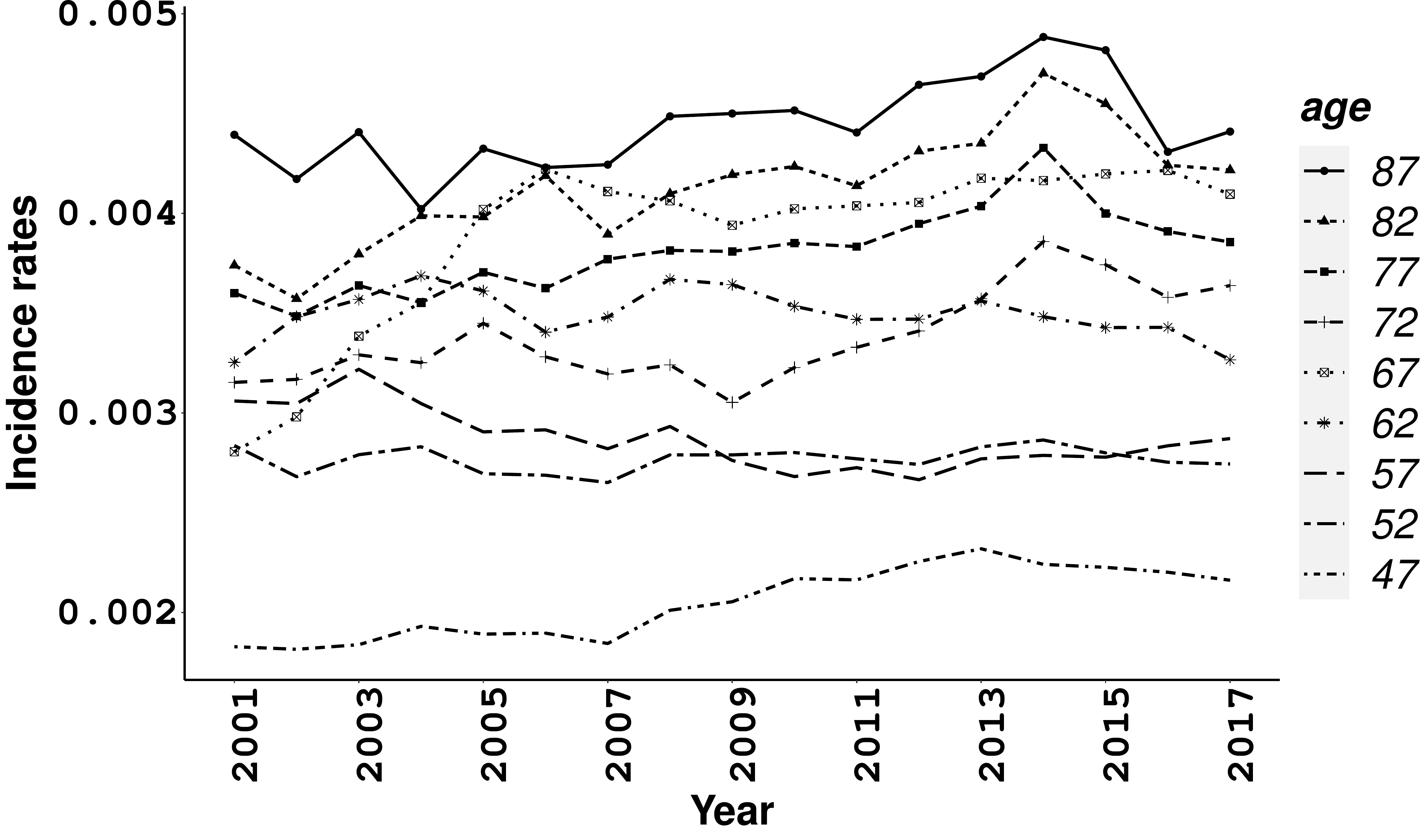}}
	\hfill
	\subfloat[Breast cancer mortality \label{fig:BreastMortality}]{\includegraphics[width=0.5\textwidth]{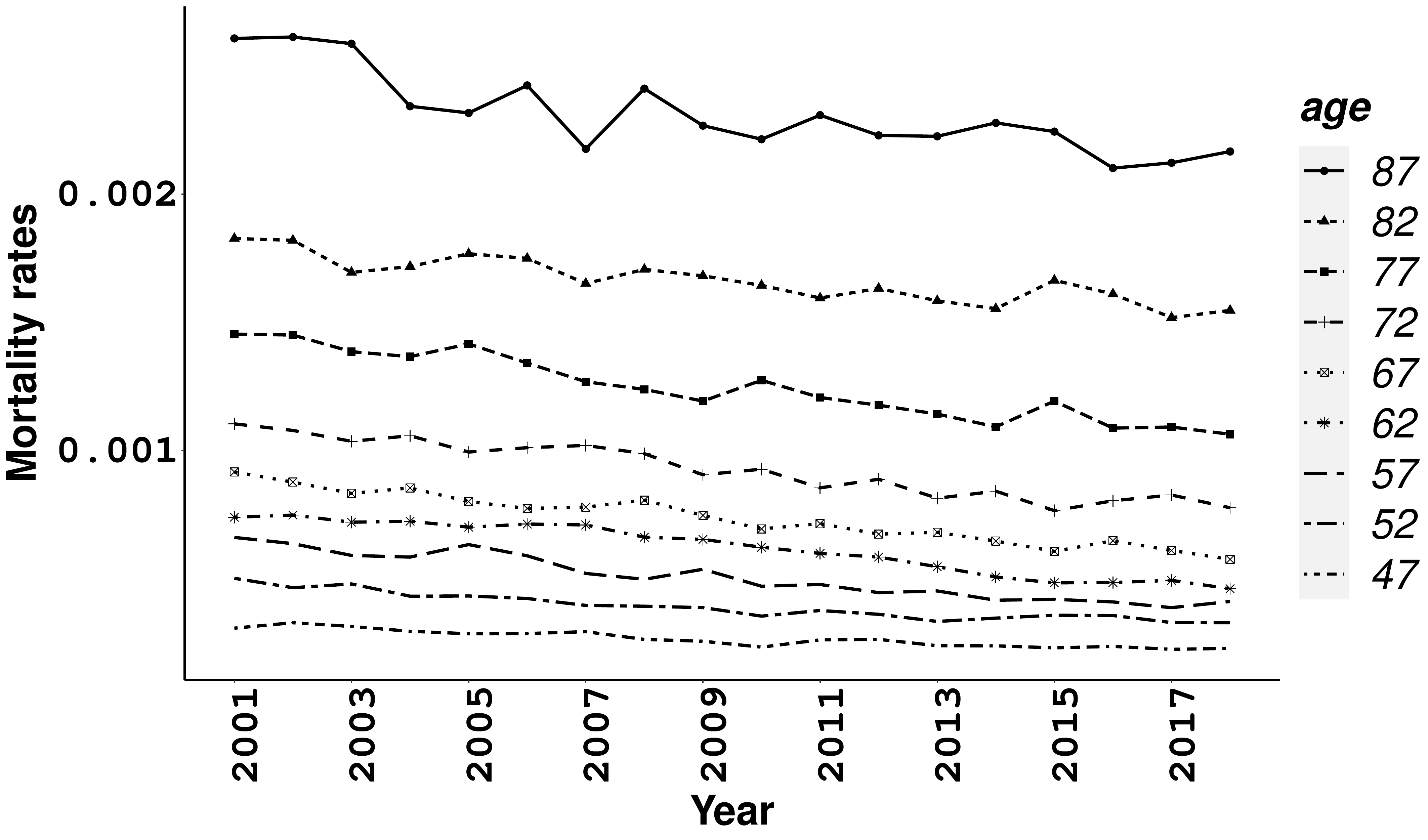}}
	\hfill
	\subfloat[Mortality from other causes (except breast cancer) \label{fig:OtherMortality}]{\includegraphics[width=0.5\textwidth]{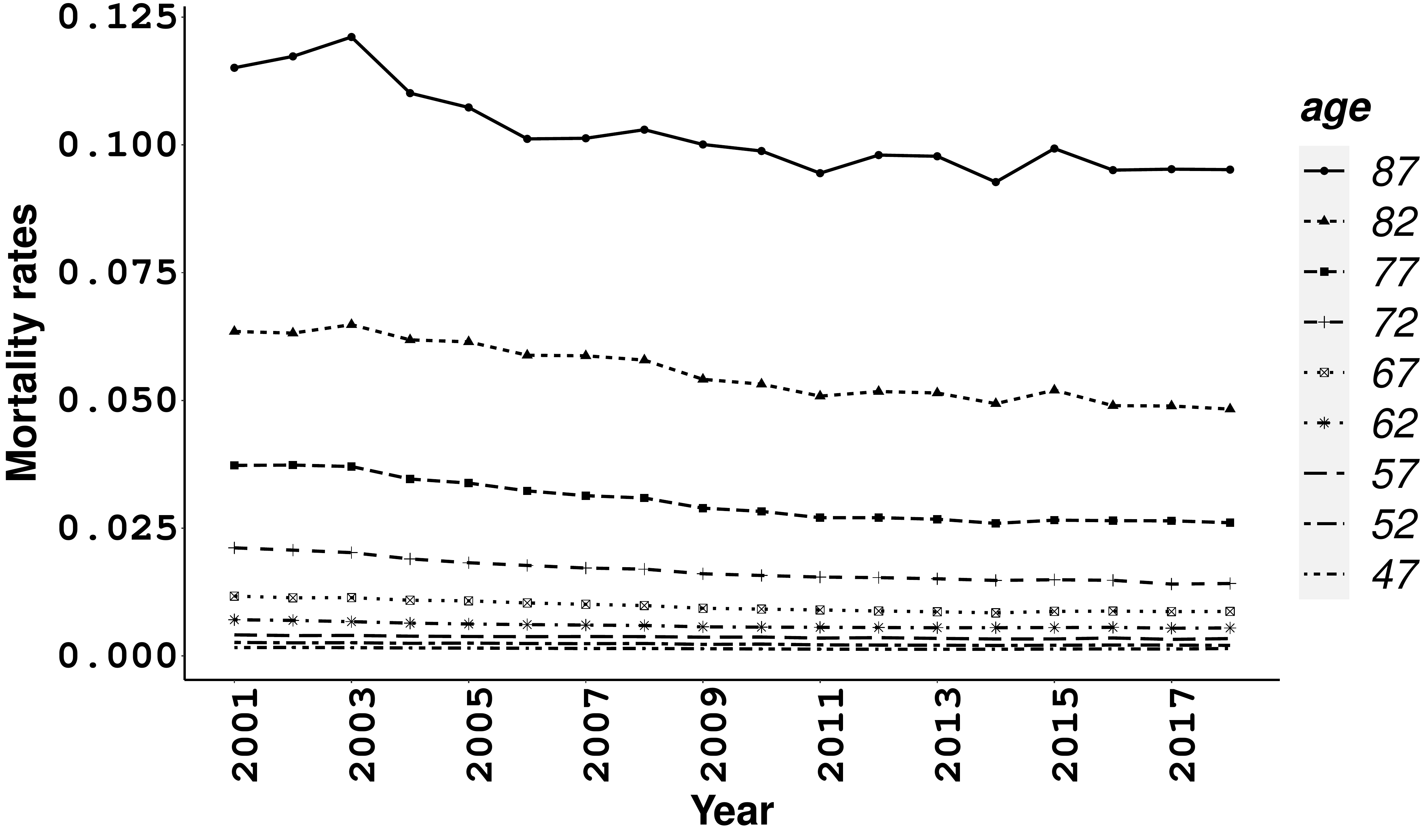}}
	\caption{Breast cancer incidence, mortality, and all-cause mortality (excluding breast cancer).}
	\label{fig:ONSData}
	\floatfoot{Note: The data is by five-year age groups between 2001--2017/2018 in England.
}
\end{figure}

\figref{fig:ONSData} exhibits available ONS-provided data at various ages, including screening age groups 47--73, {from 2001 to 2017 for cancer incidence and up to 2018 for mortality.
Note that the first screening programme was introduced in 1988, targeting women aged 50--64. 
Later, the screening was extended} to age 70 between 2002 and 2004, {including the age groups 47--73 at which screening takes place since an announcement made in 2007} \citep{Quinnetal1995, BreastScreening, Duffyetal2010, NHS2021BreastCancerExtension}. 
In \figref{fig:ONSData}, five-year age groups are represented by their mid-points. 
\figref{fig:BreastIncidence} shows BC incidence, which is calculated as new cancer registrations divided by mid-year population estimates, and generally shows an increasing trend over calendar time at all ages with higher incidence at older ages.
\figref{fig:BreastMortality} shows BC mortality, which is calculated as deaths from BC divided by mid-year population estimates, and points out a decreasing trend. 
Mortality from other causes, not including BC as a cause, shows a more heterogeneous distribution across different ages with a decreasing trend 
(\figref{fig:OtherMortality}). 

{\subsection{Key transition intensities}

For obtaining the transition intensities in \figref{fig:BC_model}, and 
in the absence of a large-scale study covering all necessary transitions, we determine the key transition intensities 
based on available data and published studies, as shown in \tabref{tab:TransitionIntensities1.v4}. 
What follows in this section summarises sources that we have used to calibrate the overall process.
}
\begin{table}[H]
	\centering
	\caption{Age-specific transition intensities for the semi-Markov model in \figref{fig:BC_model}.}
	\stackunder{
	\begin{tabular}{rrrrr}
		\hline
		Age &  $\mu^{01}_x$ & $\mu^{04}_x$  & $\mu^{35}_x$  \\
		\hline
		65--69 &0.00333  &0.00878 & 0.28060\\ 
		70--74 & 0.00286 &0.01521  & 0.36002 \\ 
		75--79 & 0.00324 &0.02693 & {0.40000} \\ 
		80--84 & 0.00355  &0.05142 & 0.49711 \\ 
		85--89 & 0.00377 &0.09684 & { 0.50000} \\ 
		\hline
	\end{tabular}\label{tab:TransitionIntensities1.v4}
}
{\parbox{3.5in}{
		\footnotesize Note: $\mu^{01}_x$ and $\mu^{04}_x$ are based on the ONS data, $\mu^{35}_x$ is based on a published study.\\
		\footnotesize Source: See Section~\ref{sec:ONSData} and \citet{Zhouetal2020}.}
}
\end{table}

{For simplicity, we assume that transition intensities to death due to other causes from all `live' states 
are equal to each other, particularly equal to $\mu^{04}_x$, shown as follows:}
\begin{eqnarray}
\mu^{14}_x = \mu^{24}_x = \mu^{34}_x = \mu^{04}_x.
\end{eqnarray}

The transition intensity from State 0 to State 4, $\mu^{04}_x$, are determined using deaths from other causes from 2010 to 2015 in England, divided by the corresponding mid-year population estimates in the same years (see Section~\ref{sec:ONSData}). {The time period is chosen so that it is consistent with the time period of other transition intensities.

Note that we ignore any time trend in BC incidence and mortality rates, or in mortality rates from other causes, in the calculation period (1 January 2020 to 31 December 2024). Also, to the best of our knowledge, there is no available literature regarding the level of $\alpha$ and $\beta$, which are used to determine $\mu^{02}_x$ and $\mu^{23}_{x,z}$, respectively, in \eqref{eq:Parametrisation}. 
We consider a range of values between 0.4 and 0.8 for $\alpha$ and between $\frac{1}{5}$ and $\frac{1}{10}$ for $\beta$. 
}


\subsubsection{Determining $\mu^{01}_x$: Clinical diagnosis of breast cancer}

{We do not have suitable empirical data on clinical diagnosis by age and stage. This is particularly important for determining transitions to State 1 and State 3. Available data include 
BC registrations by stage in England for year of diagnosis 2012--2015 \citep{ONSData2016}. 
However, it is not recommended to use this yearly information \citep{ONS2016CancerSurvival}, due to issues relating to the potential incomplete nature of the data. Therefore, we determine the transition intensities $\mu^{01}_x$, based on 81\% of overall cancer registrations, provided by the ONS, as suggested in \citet{ONS2016CancerSurvival} (Table 1). For consistency with the $\mu^{35}_x$ intensities, which were obtained based on data between 2010--2015 (see Section~\ref{sec:Datamu35}), we also determine $\mu^{01}_x$ based on the same time period. 
The ONS mid-year population estimates for England, during the same time period, are used to calculate the exposure in State 0. 
The resulting transition intensities $\mu^{01}_x$ are shown in \tabref{tab:TransitionIntensities1.v4}, along with other key transition intensities.}

We note that an alternative source for defining $\mu^{01}_x$ could be cancer registrations reported by \citet{Rutherfordetal2013, Rutherfordetal2015} (See Table 1 in \citet{Rutherfordetal2013, Rutherfordetal2015}). 
Although this data is more granular than the ONS data, stratified by both age and stage for women in the east of England between 2006--2010, the corresponding exposure is not available from the same source. 
Therefore, we have chosen to use the ONS data for our results. 

\subsubsection{{Determining $\mu^{13}_x$: Developing metastatic breast cancer}}\label{sec:Determiningmu13}

\begin{figure}[H]
	\centering
	\subfloat{\includegraphics[width=0.5\textwidth]{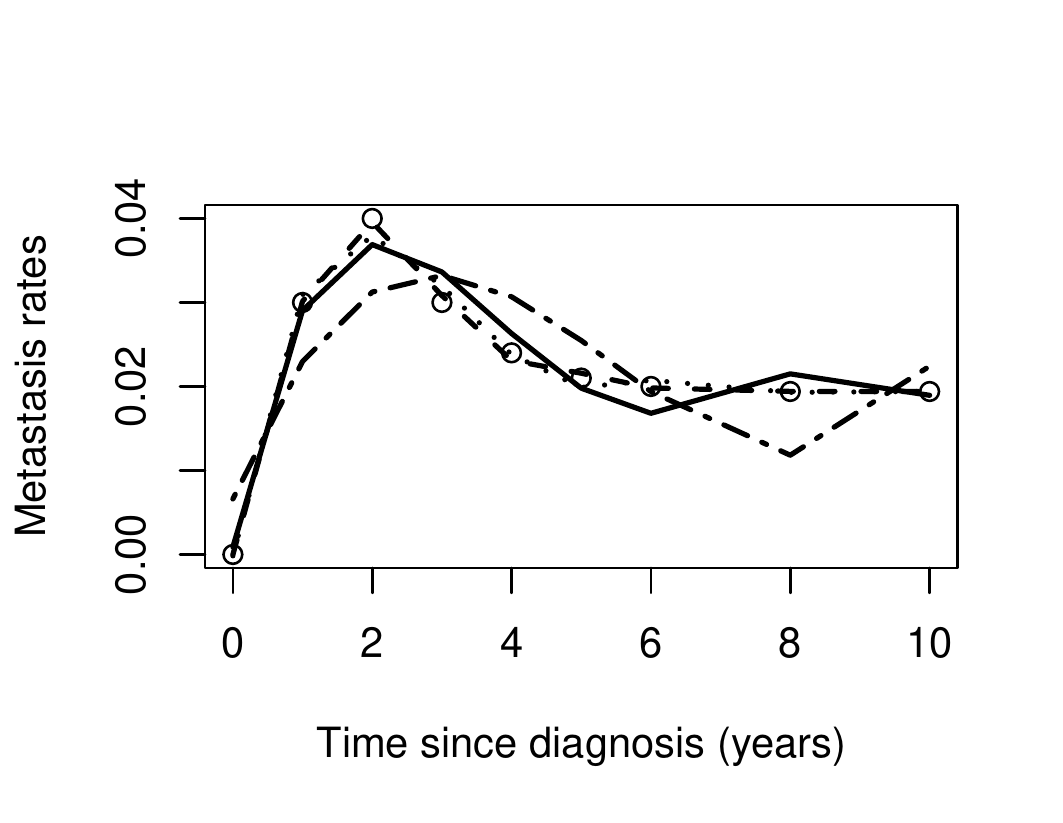}}
	\caption{Rates of transition from State 1 to State 3.
	} 	\label{fig:ColzaniFig1}
\floatfoot{Note: Circles are observed values and lines are fitted values from $3^{\text{rd}}$ (two dash), $4^{\text{th}}$ (solid), $6^{\text{th}}$ (dotted), and $7^{\text{th}}$ (dashed) degree polynomials.\\
Source: Figure 1 in \citet{Colzanietal2014}.}
\end{figure}

\citet{Colzanietal2014} {estimate risk of developing first distant metastasis} by age within 10 years of diagnosis of first invasive BC for women in Stockholm and Gotland Swedish counties between 1990--2006, noting fairly stable rates after a peak at about 2 years for women older than 50 years ({\figref{fig:ColzaniFig1}}). 

We assume that the transition intensity from State 1 to State 3, $\mu^{13}_{x,z}$, follows a functional form, indicating a steep increase in the first two years with stable rates afterwards, based on  \citet{Colzanietal2014}. \figref{fig:ColzaniFig1} shows observed values, taken from Figure 1 in \citet{Colzanietal2014}, 
and fitted values based on some polynomial functions.

Here, we {define $\mu^{13}_{x, z}$ as a function of duration only}, using a $4^{\text{th}}$ degree polynomial, given as 
\begin{equation}\label{eq:PolyFuncMu13}
\mu^{13}_{x, z} = 0.00088644 + 0.04191138 z - 0.01574062 z^2 + 0.00207282 z^3 - 0.00008998 z^4,
\end{equation}
for {a given age $x$ }and $0 \le z < 10$. 
This function is not suitable for extrapolation to durations $z >10$.
Parameters are estimated from the data in \tabref{tab:ColzaniData}.

\begin{table}[H]
	\centering
	\caption{Rates of transition from State 1 to State 3 in different {durations (years)}.
	}\label{tab:ColzaniData}
	\stackunder{
	\begin{tabular}{lccccccccc}
		\hline
		{time}             & 0 & 1    & 2    & 3    & 4     & 5     & 6    & 8      & 10     \\
		{$\mu^{13}_{x,z}$} & 0 & 0.03 & 0.04 & 0.03 & 0.024 & 0.021 & 0.02 & 0.0194 & 0.0194 \\ \hline
	\end{tabular}
}
{\parbox{4in}{
		\footnotesize Note: The values are determined based on Figure 1 in \citet{Colzanietal2014}.
}
}
\end{table}

{We also consider a special case of the semi-Markov model in \figref{fig:BC_model}, assuming $\mu^{13}_x=0.01954$. This value represents average of first distant metastasis rates based on Table 1 in \citet{Colzanietal2014}. 
	Note that rates of transition from State 2 to State 3, $\mu^{23}_x$, are determined based on $\mu^{13}_x=0.01954$, following \eqref{eq:Parametrisation}.
}

\subsubsection{{Determining $\beta$: Measure of treatment effectiveness}} \label{sec:TumourGrowthBC}

State 2 is important in our model for being able to quantify the potential impact of a major disruption to health services on cancer mortality. However, there is no empirical data regarding unobserved BC. 
For modelling purposes we assume that rates of transition from States 1 and 2 to State 3 are related through parameter $\beta$, which represents a measure of treatment effectiveness, as shown in \eqref{eq:Parametrisation}. 
There is no available data regarding how a BC tumour can grow in the absence of treatment, 
although this is expected to differ by tumour subtypes.
This is mainly because patients are required to be treated as soon as they are diagnosed \citep{Nakashimaetal2018}. 
However, there is information in the literature about tumour growth for patients waiting for surgery {that can be used as a proxy for the tumour growth in the lack of treatment leading to a more advanced BC stage}. 
We use this to establish a reasonable value for $\beta$.

\citet{Leeetal2014} quantify tumour growth rates for 1328 women diagnosed with invasive BC, during wait times for surgery, at Seoul National University Hospital between 2013--2014. 
They report significant changes depending on surrogate molecular subtypes, e.g. larger diameter changes in more aggressive molecular subtypes, and a frequent upgrade from Stage 1 to Stage 2 during waiting times for surgery, where the median waiting time is 31 days. 
\citet{Nakashimaetal2018} report significant changes in tumours between diagnosis and surgery for 64\% of 309 patients diagnosed with invasive BC between 2014--2016, where the mean waiting time is 56.9 days. 
\citet{Yooetal2015} report significant increases in tumour sizes of 55\% of 957 patients, diagnosed with invasive BC between 2002--2010, where the median time interval between initial and second examination is 28 days.  
This information suggests a considerable change in BC tumours for more than half of the observed populations during a period of one or two months, and therefore points towards the transition intensity
$\mu_{x,z}^{23}$ being considerably higher than $\mu_{x, z}^{13}$, in the absence of any treatment. 
{We consider a range of values between $\frac{1}{5}$ and $\frac{1}{10}$ for $\beta$ in the absence of empirical data and literature information.}

\subsubsection{{Determining $\mu^{35}_x$: Metastatic breast cancer related mortality}} \label{sec:Datamu35}
Survival from metastatic BC can be highly correlated to age, tumour type, and treatment, in addition to other patient- or disease-related factors \citep{denBroketal2017, Purushothametal2014}.
\citet{Zhaoetal2020} report BC deaths by age within 12 months of Stage 4 BC diagnosis, using a cohort, between 2010--2015, obtained from the National Cancer Institute Surveillance, Epidemiology and End Results (SEER) database. 
We define rates of transition to State 5, $\mu^{35}_x$, based on the numbers shown in Table 1 in \citet{Zhaoetal2020}. 
	Note that `No early death' shows the number of patients that survived for 12 months, whilst `Total early death' displays the number of patients deceased within 12 months in that study. Thus, we {use} a Uniform Distribution of Deaths assumption, to define the exposure under `Total early death' \citep{Hossain1994}.
{Specifically, we assume that `No Early Death' contributes a full year and each `Early Death' half a year on average to the exposure.}
The resulting rates, $\mu^{35}_x$, presented in \tabref{tab:TransitionIntensities1.v4}, are assumed to remain unchanged during {the calculation period from 1 January 2020 to 31 December 2024.
Note that we add small increments to the rates at ages 75--79 and 85--89 where these are rounded to 0.4 and 0.5, respectively.
}

\section{{Pandemic} Scenarios}  \label{secb:Results}

We consider two pandemic scenarios. Scenario 1 (S1) introduces a significant change in transitions to death from other causes, but does not involve any BC-related assumption. 
{Thus, it reflects what would have been expected if the pandemic-related health disruptions had not affected BC diagnosis.}
In Scenario 2 (S2) we additionally assume a decline in cancer diagnoses.

\begin{itemize}
	\item[{\bf S1}:] The pandemic is assumed to result in increased deaths from other causes. 
	This accords with empirical evidence (Section~\ref{Sec:Assumption_ExcessMortality}).
	\item[{\bf S2}:] In addition to the assumption in S1, we further assume a decline in BC diagnosis, i.e. a decline in the number of transfers to State 1 (Section~\ref{Sec:Assumption_DiagnosisDecline}). 
	This is represented by {changing the level of a given $\alpha$ in \eqref{eq:Parametrisation} based on \tabref{tab:Summary_assumptions}}.
	Since we assume that the onset of BC remains unchanged before and after the pandemic, see \eqref{eq:BCDiagUnchnaged} and \eqref{eq:Parametrisation}, 
	we accordingly adjust the total transition intensity into State 2, $\mu_x^{02}$ (Assumption A4). 
\end{itemize}

{\tabref{tab:Summary_assumptions} summarises the assumptions made in relation to some of the key transition intensities in the pandemic scenarios. These assumptions are explained in Sections~\ref{Sec:Assumption_ExcessMortality}--\ref{Sec:Assumption_DiagnosisDecline}.

\begin{table}[H]
	\caption{Proportionality constants applied to transition intensities in the pandemic scenarios.}
		\label{tab:Summary_assumptions}
\stackunder{	\begin{tabular}{lccc}
		\hline
		\multicolumn{1}{c}{Pandemic period} & $\mu^{01}_x/ \mu^{02}_x$ & \multicolumn{2}{c}{$\mu^{04}_x$} \\ \hline
		\multicolumn{1}{c}{}                &       $\alpha$         & 65--84          & 85--89         \\ \hline
		April--Nov. 2020                    & 0.8                & 1.13            & 1.12           \\
		Dec. 2020--Nov. 2021                & 1                    & 1.13            & 1.12           \\
		Dec. 2021--Dec. 2022                & 1                  & 1.10            & 1.09           \\
		Jan.--Dec. 2023                     & 1                 & 1.07            & 1.06           \\
		Jan.--Dec. 2024                     & 1                 & 1.04            & 1.03           \\ \hline
	\end{tabular}}
{\parbox{3in}{
		\footnotesize Note: Proportionality constants are the same across all ages in both pandemic scenarios.
	}
}
\end{table}}

\subsection{Scenario 1: Excess mortality due to COVID-19 in England} \label{Sec:Assumption_ExcessMortality}

{There is evidence suggesting that the COVID-19 pandemic has caused an increase in excess mortality, which can be linked to Scenario 1.  }
The Office for Health Improvement and Disparities (OHID) in England monitors excess mortality by age, sex, Upper Tier Local Authority, ethnic group, level of deprivation, cause of death and place of death since 21 March 2020, in order to have a better understanding of the impact of COVID-19. 
They report ratios representing relative changes between registered and expected excess deaths for each group \citep{OfficeforHealthImpDis2022}. 
We use {a set of ratios, shown in \tabref{tab:Summary_assumptions}, to define} the potential increase in transition to death from other causes.

The age-specific transition intensities to death due to other causes, $\mu^{04}_x$, are assumed to increase by a factor of 1.13 for ages 65--84 and 1.12 for ages 85+ from April 2020 until November 2021, while we assume they increase by a factor 1.10 for ages 65--84 and 1.09 for ages 85+ from November 2021 until the end of 2022 \citep{OfficeforHealthImpDis2022}. 
Given the gradual {decrease in the excess mortality} between April 2020 and December 2022, 
we assume that $\mu^{04}_x$ could still be higher than the pre-pandemic levels for an additional period of  two years. 
Specifically, $\mu^{04}_x$ is assumed to increase by the following factors: 1.07 for ages 65--84 and 1.06 for ages 85+ in 2023;  
1.04 for ages 65--84 and 1.03 for ages 85+ in 2024.

\subsection{Scenario 2: Changes in breast cancer risk amid COVID-19}\label{Sec:Assumption_DiagnosisDecline}

There is no evidence suggesting that the COVID-19 pandemic increased BC incidence. 
Therefore, we assume that overall new cases of cancer are not affected by the pandemic (A4 under Section~\ref{Sec:BCModelAssumptions}). 
This implies that the onset of BC is assumed to be unchanged by the pandemic, and therefore $\mu^{*}_x$ is not affected.
We further assume that there is no time trend in BC risk over the next five years.

However, cancer registrations are known to have reduced during national lockdowns \citep{CRUKCIT2021}. 
Particularly, Public Health Scotland (PHS) reported that BC registrations were 19\% lower than the 2018/2019 average during the nine months of the pandemic (April--December 2020), as a result of initial health disruptions \citep{PHS2021}. 
The number of BC registrations in the second quarter of 2020 is noted to start returning back to the pre-pandemic levels towards the end of 2020.
Based on the available information, we assume that, for all ages, diagnosis of BC, $\mu^{01}_x$, is decreased by 20\% from April 2020 until the end of 2020. Following that, it is then assumed that they are restored back to  pre-pandemic levels. 	
The intensity $\mu_x^{02}$ is adjusted accordingly, keeping the overall BC onset rate unchanged {(see \eqref{eq:Parametrisation} and \tabref{tab:Summary_assumptions}).}

\section{Results}  \label{secc:Results}

{In this section we present the main findings based on different scenarios, associated with  a pre-pandemic model calibration and pandemic scenarios S1 and S2. 
These findings are obtained for selected values of $\alpha$ and $\beta$, which are $\alpha=0.6\,\, \text{and}\,\, \beta=\frac{1}{7}$.
We note the lack of data to determine the values of these parameters. 
Therefore, we test for sensitivity of the results to changes in  $\alpha$ and $\beta$ in Section~\ref{Sec:SensitivityAnalysis}.} 

\tabref{tab:TransitionNumbersSurvRatesbyAge} compares age-specific occupancy probabilities, denoted by ${_{t}} p^{ij}_x$ from state $i$ to state $j$ at age $x$, {based on the semi-Markov BC model, \figref{fig:BC_model}, over one and 5 years from 1 January 2020. 
	As a special case of the model in \figref{fig:BC_model}, we also present results with a Markov model, which is determined by removing duration dependency in rates of transition from State 1 to State 3, $\mu^{13}_{x, z}$, and accordingly in $\mu^{23}_{x, z}$, as well. Therefore, 
	in the Markov model, we determine constant values for $\mu^{13}_x$ and $\mu^{23}_x$ over both age and time (Section~\ref{sec:Determiningmu13}). 
	This simplification can additionally allow us to compare results from the Markov and semi-Markov models.
}

	\begin{landscape}
		\begin{table}[H]
			\centering
			\caption{Occupancy probabilities for women (\%) being in different states over 5 years.}
			\label{tab:TransitionNumbersSurvRatesbyAge} 	
			\stackunder{
				\begin{tabular}{lrr>{\em}rr>{\em}rr>{\em}rrr>{\em}rr>{\em}rr>{\em}rrr} 
					\hline
					&                \multicolumn{16}{c}{{Occupancy Probabilities (\%)}} 
					\\ \hline
					&                \multicolumn{10}{c}{{From State 0}} &   \multicolumn{4}{c} {From State 1} &   \multicolumn{2}{c}{From State 3}\\ \hline
					Age & $  {_{5}} p^{00}_x $& \multicolumn{2}{c}{ ${_{5}} p^{01}_x $}  &\multicolumn{2}{c}{ ${_{5}} p^{02}_x $} & \multicolumn{2}{c}{ ${_{5}} p^{03}_x $}& $ {_{5}} p^{04}_x $ &\multicolumn{2}{c}{ ${_{5}} p^{05}_x  $}  & \multicolumn{2}{c}{ ${_{1}} p^{15}_x  $} & \multicolumn{2}{c}{ ${_{5}} p^{15}_x  $} & $ {_{1}} p^{35}_x $ & $ {_{5}} p^{35}_x $ \\ 
					\hline
					&\multicolumn{1}{c} {M}  & \multicolumn{1}{c} {M} & \multicolumn{1}{c} {S-M} & \multicolumn{1}{c} {M} & \multicolumn{1}{c} {S-M} & \multicolumn{1}{c} {M} & \multicolumn{1}{c} {S-M} & \multicolumn{1}{c} {M} & \multicolumn{1}{c} {M} & \multicolumn{1}{c} {S-M} & \multicolumn{1}{c} {M} & \multicolumn{1}{c} {S-M}&
					\multicolumn{1}{c} {M} & \multicolumn{1}{c} {S-M} & \multicolumn{1}{c} {M}& \multicolumn{1}{c} {M} \\
					\hline
					& \multicolumn{16}{c} {Pre-pandemic calibration} \\
					65--69 & 93.09 & 1.50 & 1.47 & 0.76 & 0.68 & 0.24 & 0.31 & 4.29 & 0.13 & 0.16 & 0.25 & 0.16 & 4.24 & 5.98 & 24.36 & 74.15 \\ 
					70--74 & 90.49 & 1.25 & 1.22 & 0.63 & 0.57 & 0.18 & 0.23 & 7.32 & 0.13 & 0.16 & 0.31 & 0.20 & 4.82 & 6.82 & 30.02 & 81.25 \\ 
					75--79 & 85.07 & 1.33 & 1.31 & 0.67 & 0.61 & 0.18 & 0.24 & 12.59 & 0.15 & 0.19 & 0.34 & 0.22 & 4.92 & 6.97 & 32.56 & 82.61 \\ 
					80--84 & 75.07 & 1.29 & 1.26 & 0.65 & 0.59 & 0.15 & 0.20 & 22.66 & 0.17 & 0.21 & 0.40 & 0.26 & 5.09 & 7.21 & 38.26 & 84.79 \\ 
					85--89 & 59.71 & 1.09 & 1.07 & 0.55 & 0.50 & 0.13 & 0.17 & 38.36 & 0.16 & 0.19 & 0.39 & 0.25 & 4.47 & 6.29 & 37.65 & 79.54 \\  
					& \multicolumn{16}{c} {Pandemic scenarios} \\
					S1 & & & & & & & & & & & & & & & & \\
					65--69 & 92.73 & 1.49 & 1.46 & 0.75 & 0.68 & 0.23 & 0.31 & 4.66 & 0.13 & 0.16 & 0.25 & 0.16 & 4.23 & 5.96 & 24.36 & 74.03 \\ 
					70--74 & 89.90 & 1.24 & 1.22 & 0.63 & 0.56 & 0.18 & 0.23 & 7.93 & 0.13 & 0.16 & 0.31 & 0.20 & 4.80 & 6.79 & 30.00 & 81.03 \\ 
					75--79 & 84.09 & 1.32 & 1.29 & 0.67 & 0.60 & 0.18 & 0.23 & 13.60 & 0.15 & 0.19 & 0.33 & 0.22 & 4.88 & 6.91 & 32.53 & 82.24 \\ 
					80--84 & 73.42 & 1.26 & 1.24 & 0.64 & 0.57 & 0.15 & 0.20 & 24.36 & 0.17 & 0.21 & 0.40 & 0.26 & 5.02 & 7.10 & 38.20 & 84.15 \\ 
					85--89 & 57.53 & 1.05 & 1.03 & 0.53 & 0.48 & 0.13 & 0.17 & 40.61 & 0.15 & 0.19 & 0.39 & 0.25 & 4.36 & 6.12 & 37.55 & 78.56 \\ 
					S2& & & & & & & & & & & & & & & & \\
					65--69 & 92.73 & 1.45 & 1.42 & 0.78 & 0.70 & 0.24 & 0.32 & 4.66 & 0.14 & 0.17 & 0.25 & 0.16 & 4.23 & 5.96 & 24.36 & 74.03 \\ 
					70--74 & 89.90 & 1.20 & 1.18 & 0.65 & 0.58 & 0.18 & 0.24 & 7.93 & 0.14 & 0.17 & 0.31 & 0.20 & 4.80 & 6.79 & 30.00 & 81.03 \\ 
					75--79 & 84.09 & 1.28 & 1.25 & 0.69 & 0.62 & 0.18 & 0.24 & 13.60 & 0.16 & 0.20 & 0.33 & 0.22 & 4.88 & 6.91 & 32.53 & 82.24 \\ 
					80--84 & 73.42 & 1.22 & 1.20 & 0.66 & 0.59 & 0.16 & 0.21 & 24.36 & 0.18 & 0.22 & 0.40 & 0.26 & 5.02 & 7.10 & 38.20 & 84.15 \\ 
					85--89 & 57.53 & 1.02 & 1.00 & 0.55 & 0.49 & 0.13 & 0.17 & 40.61 & 0.16 & 0.20 & 0.39 & 0.25 & 4.36 & 6.12 & 37.55 & 78.56 \\ 
					\hline
			\end{tabular}}
			{\parbox{7in}{
					\footnotesize Note: Women have no breast cancer or clinically diagnosed with breast cancer at time zero. Results are based on Markov (M) and semi-Markov (S-M) models in the pre-pandemic model calibration and the pandemic scenarios, Scenario 1 (S1) and Scenario 2 (S2), for {$\alpha=0.6$, $\mu^{13} = \frac{1}{7}\mu^{23}$.} }
			}
		\end{table}
		
	\end{landscape}

\subsection{{Unobserved and observed breast cancer cases}}
\mbox{} \\

\tabref{tab:TransitionNumbersSurvRatesbyAge} shows that for a woman free of BC at time zero, the probability of being diagnosed with pre-metastatic BC over the following 5 years, ${_{5}} p^{01}_x$, has decreased {by 3--6\%, across different ages, in Scenario 2, as compared to the pre-pandemic calibration.} The results show bigger changes at older ages in Scenario 2, consistent in both models. The decline in ${_{5}} p^{01}_x$ has remained less than 3\% in Scenario 1. {At the same time, the probability of having BC and staying undiagnosed, ${_{5}} p^{02}_x$, increases by 1--3\% over 5 years in Scenario 2 based on the Markov model. 
The increase is mostly higher at younger ages. 
The increase in the same probability, ${_{5}} p^{02}_x$,  is less than 2\%  under the semi-Markov model.}

Meanwhile, results under both models show that for a woman with no BC at time zero, the probability of being diagnosed with metastatic BC over the following 5 years, 
${_{5}} p^{03}_x$, increases at certain ages, {for instance, by 5\% to 6\% at ages 80--84 in Scenario 2, as compared to the pre-pandemic calibration}. An increase in ${_{5}} p^{03}_x$, up to 4\%, occurs at ages 65--69 and 70--74 based on the semi-Markov model. 

{In Scenario 1 the modelling mostly reveals a decline in ${_{5}} p^{02}_x$, up to 3\%, as compared to the pre-pandemic levels based on both models, and no considerable changes in ${_{5}} p^{03}_x$, apart from the decrease in the youngest age in the Markov model.}
The decrease in ${_{5}} p^{02}_x$ and occasional decrease in ${_{5}} p^{03}_x$ {in Scenario 1} can be associated with the increase in deaths from other causes, 
since the transition intensities from States 2--3 to `Dead, Other Causes' are assumed to be equal to $\mu^{04}_x$.

These findings are aligned with documented information that cancer patients have been more vulnerable to the SARS-CoV-2 coronavirus and affected worse by the pandemic, compared to the general population \citep{Pinatoetal2020, Garassinotal2020, Leeetal2020, Sainietal2020}. It is also worth noting that the PHS reported falls in Stages 1--2 BC in Scotland along with small increases in Stages 3--4 BC in 2020 \citep{PHS2021}.

\subsection{{Breast cancer mortality}}
\mbox{} \\
For women with clinical cancer diagnosis, {i.e. women in either  State 1 or State 3 at time zero}, we define cancer mortality as the probability of moving to State 5, 
for the period under consideration. 

The dependence of BC mortality on age becomes more evident if we consider a longer period after diagnosis, where bigger changes are observed in more advanced ages {for women with metastatic BC,} consistent in both models (\tabref{tab:TransitionNumbersSurvRatesbyAge}). 
For instance, in the pre-pandemic calibration, one-year mortality for a woman aged 65--69 with metastatic BC is estimated as 24.36\%, whereas at ages 80+ one-year mortality {is} around and above 37\%.
On the other hand, {the variation in mortality, with respect to different ages, for women in State 1 is very small} even after 5 years.

The results in \tabref{tab:TransitionNumbersSurvRatesbyAge} also show that 
mortality in {5 years} after metastatic BC diagnosis is estimated to be between 74.15--84.79\%, 
whereas the mortality for a woman with pre-metastatic BC diagnosis differs in the presence of duration dependence: \begin{enumerate*}[label=(\roman*)]
	\item around 4--5\% under the Markov model; and
	\item 6--7\% under the semi-Markov model.
\end{enumerate*}
Meanwhile, the relationship between 5-year mortality and age is not straightforward to interpret due to the following reasons: 
\begin{itemize}
	\item We have simplified BC progression using two states, with BC Stages 1--3 being combined and included in States 1 and 2, due to lack of reliable data. Ideally, BC Stage 3, which indicates locally advanced BC, should be treated differently than Stages 1 and 2, since  survival from Stage 3 can be markedly different than that from Stages 1 and 2 \citep{Rutherfordetal2015, Maringeetal2020}. 
	\item In the absence of sufficient data, we have assumed constant transition intensities over periods of 5 years. Given the trends of BC incidence and mortality over time in \figref{fig:ONSData}, this may not be realistic.
	\item The probability of metastasis decreases with age, while mortality risk increases with age in the presence of any BC-related condition \citep{Purushothametal2014}. 
	The net effect of these two forces might be another reason for not seeing a consistent trend by age in 5-year BC mortality rates.
\end{itemize}

All-cause mortality, {including death from BC}, for women with pre-metastatic or metastatic BC is also presented over periods of 5 years, where age dependence is clear (\tabref{tab:TransitionsbyAge1year} and \tabref{tab:TransitionsbyAge1year_CovidScenarios}).

{There is a relative decline in the cancer mortality, less than 2\%, across different ages in the pandemic scenarios in comparison to the pre-pandemic calibration under both models. 
	This decline is as a result of increases in excess mortality (Section~\ref{Sec:Assumption_ExcessMortality}).} However, across pandemic scenarios, our modelling shows no change in the cancer mortality for women with clinical diagnoses (\tabref{tab:TransitionNumbersSurvRatesbyAge}). This is because our approach assumes that there is no change in the onset of BC before and after the pandemic, and the corresponding probabilities are conditional on BC diagnosis.

{The models also allow us to obtain cancer survival rates. }
Cancer-specific survival, as used by the ONS, is one of most widely accepted survival measures.
It is stated to be a {`net'} measure and interpreted as the number of people being alive {`after cancer diagnosis'}. This measure is considered to represent a {`hypothetical situation in which the cancer of interest is the only possible cause of death'} \citep{Mariottoetal2014, SwaminathanandBrenner2011, ONSSurv2019}. 
{We refer to this as the `ONS approach'.}
For a woman diagnosed with pre-metastatic BC at age $x$, for instance, cancer-specific survival in $t$ years can be obtained {based on the ONS approach} as follows:
\begin{equation}
\frac{1 -  {_{t}} p^{14}_x  \, - {_{t}} p^{15}_x \, }{1 - {_{t}} p^{14}_x  },
\label{eq:CancerSurvival}
\end{equation}
where ${_{t}} p^{14}_x$  represents mortality from other causes, while ${_{t}} p^{15}_x$ represents mortality from BC.

\tabref{tab:CancerSpecificSurvRatesbyAge} compares 1-, 5-, and 10-year survival probabilities based on both Markov and semi-Markov models in the pre-pandemic calibration {using \eqref{eq:CancerSurvival} based on the ONS approach and an adjustment of our models. 
	We adjust the models developed here by setting the transition intensities} to `Dead, Other Causes' {after being diagnosed with BC or having BC without a clinical diagnosis, i.e. $\mu_x^{14},\, \mu_x^{24}$ and $\mu_x^{34}$, equal to zero.} 
This allows `Dead, BC' to be the only cause of death.


	\begin{landscape}
\begin{table}[H]
\centering
\caption{{1-, 5-, and 10-year survival probabilities (\%) for women from breast cancer.}}
	\label{tab:CancerSpecificSurvRatesbyAge} 	
		\stackunder{
\begin{tabular}{ lr>{\em}rr>{\em}rr>{\em}rr>{\em}rr>{\em}rr>{\em}rrrr}
	\hline
	      &      \multicolumn{15}{c}{{Cancer Survival (\%)}}   \\
	      \hline
	  			&          \multicolumn{6}{c}{{From State 1}}  &          \multicolumn{6}{c}{{From State 2}}   &    \multicolumn{3}{c}{{From State 3}}  \\
  \hline
Age &  \multicolumn{2}{c}{{1-year}} &  \multicolumn{2}{c}{{5-year}} &  \multicolumn{2}{c}{{10-year}}  & \multicolumn{2}{c}{{1-year}} &  \multicolumn{2}{c}{{5-year}} &  \multicolumn{2}{c}{{10-year}} & 1-year& 5-year& 10-year\\ 
  \hline
    &   M & S-M & M & S-M & M & S-M & M & S-M & M & S-M & M & S-M & M & M & M   \\
    \hline
    &   \multicolumn{15}{c}{{ONS approach}}  \\
65--69 & 99.75 & 99.84 & 95.57 & 93.76 & 87.57 & 84.44 & 98.32 & 98.90 & 74.75 & 67.52 & 42.95 & 34.87 & 75.45 & 24.09 & 5.70 \\ 
70--74 & 99.69 & 99.79 & 94.81 & 92.65 & 86.06 & 82.65 & 97.90 & 98.61 & 70.62 & 62.15 & 37.67 & 29.50 & 69.60 & 15.86 & 2.44 \\ 
75--79 & 99.66 & 99.77 & 94.38 & 92.06 & 84.95 & 81.32 & 97.68 & 98.47 & 68.47 & 59.45 & 34.70 & 26.66 & 66.71 & 12.52 & 1.49 \\ 
80--84 & 99.58 & 99.72 & 93.46 & 90.75 & 82.48 & 78.35 & 97.18 & 98.13 & 63.96 & 53.89 & 29.13 & 21.58 & 60.16 & 7.06 & 0.46 \\ 
85--89 & 99.57 & 99.72 & 92.83 & 89.97 & 79.04 & 74.17 & 97.13 & 98.10 & 61.52 & 51.41 & 24.22 & 17.45 & 59.38 & 5.98 & 0.31 \\ 
    &   \multicolumn{15}{c}{{Adjusted model}}  \\
65--69 & 99.75 & 99.84 & 95.64 & 93.85 & 87.95 & 84.93 & 98.33 & 98.90 & 75.08 & 67.88 & 43.94 & 35.84 & 75.53 & 24.59 & 6.04 \\ 
70--74 & 99.69 & 99.80 & 94.95 & 92.84 & 86.81 & 83.60 & 97.91 & 98.62 & 71.26 & 62.84 & 39.40 & 31.13 & 69.77 & 16.53 & 2.73 \\ 
75--79 & 99.66 & 99.78 & 94.66 & 92.40 & 86.38 & 83.11 & 97.70 & 98.48 & 69.66 & 60.71 & 37.75 & 29.48 & 67.03 & 13.53 & 1.83 \\ 
80--84 & 99.59 & 99.73 & 94.06 & 91.53 & 85.59 & 82.23 & 97.23 & 98.16 & 66.46 & 56.46 & 34.87 & 26.70 & 60.83 & 8.33 & 0.69 \\ 
 85--89 & 99.59 & 99.73 & 94.05 & 91.50 & 85.57 & 82.21 & 97.22 & 98.15 & 66.38 & 56.35 & 34.80 & 26.64 & 60.65 & 8.21 & 0.67 \\ 
   \hline
\end{tabular}
}
{\parbox{8in}{
		\footnotesize Note: Results are for women with pre-metastatic and metastatic breast cancer, and for women with undiagnosed breast cancer based on Markov (M) and semi-Markov (S-M) models, {using  \eqref{eq:CancerSurvival}}, in the pre-pandemic calibration for $\alpha=0.6$, $\mu^{13} = \frac{1}{7}\mu^{23}$.
	}
}
\end{table}
\end{landscape}

\tabref{tab:CancerSpecificSurvRatesbyAge} shows that cancer survival is worse at older ages. 
It suggests that cancer-specific survival probabilities based on the ONS methodology applied to our data are reasonably consistent with those  based on {the adjusted models}. 
The main difference between the models has risen for women with pre-metastatic BC, {with and without a clinical diagnosis, }where lower estimates are obtained in the longer term based on the semi-Markov model. 
The estimates {across ages} change to a slightly higher degree in the longer term based on the ONS methodology as compared to the adjusted models. 
	
{We note that our findings for women with pre-metastatic and metastatic BC are broadly agreement with the ONS statistics, where 5- and 10-year age standardised survival rates (aged 15 to 99 years) for women diagnosed with BC between 2011-2015 were reported to be above 80\% and 50\%, respectively. 
Whilst very few excess deaths for women diagnosed with Stages 1--2 BC were observed, compared with general population, after the first year of diagnosis, one-year age standardised survival rate for women diagnosed with Stage 4 BC in 2015, followed up to 2016, was noted to be 65.8\% \citep{ONS2017CancerSurvival}.
Furthermore, we found that cancer survival has worsened significantly in the absence of any treatment, in State 2, as compared to those where medical treatments were available in State 1. 
For instance, 10-year cancer survival of women with pre-metastatic BC at ages 65--69 would have declined from around 84--87\% to 34--42\%, with higher rates in the Markov model, if these women could have stayed undiagnosed and taken no medical care during the 10 years.  
This is aligned with the existing medical literature \citep{Verkooijenetal2005, Josephetal2012}. 
} 

Cancer survival probabilities in the pandemic scenarios are not provided in \tabref{tab:CancerSpecificSurvRatesbyAge}. 
This is because survival is conditioned upon diagnosis of BC, which is the event disrupted by the pandemic.

\subsection{{Excess deaths}}
\mbox{} \\
The estimated numbers of deaths over 5 years, by age, due to BC and other causes, can be determined by using $ {_{5}} p^{05}_x$ and $ {_{5}} p^{04}_x$, respectively.
Estimates of excess deaths, in the corresponding period, are then calculated as the differences between estimated numbers of deaths in the pre-pandemic calibration and the pandemic scenarios (\tabref{tab:DifferencesbyAge}). 
{We note that the time trend in mortality is ignored.}

Our findings show that deaths from other causes increase by 5--8\%, with higher changes at younger ages, corresponding to 363--2,255 excess deaths, {per 100,000 women at different ages}, in Scenarios 1--2, compared to the pre-pandemic calibration under both models. 
Our model also gives a 3--6\% increase in deaths from BC across different ages in Scenario 2 based on both settings, with higher increases for younger ages. 
This corresponds to 5--8 excess BC deaths at different ages under the Markov model, and 6--10 excess deaths under the semi-Markov model.

\begin{table}[H]
	\centering
\caption{
	{ Age-specific excess number of deaths and years of life expectancy lost (YLL), per 100,000 women.} 
		\label{tab:DifferencesbyAge}} 
\stackunder{
	\begin{tabular}{lr>{\em}rr>{\em}rr>{\em}rr>{\em}r}
  \hline
  & \multicolumn{4}{c}{Excess deaths} &  \multicolumn{4}{c}{YLL} \\ \hline
 Age &  \multicolumn{2}{c}{ Dead (Other) } &  \multicolumn{2}{c}{ Dead (BC)}  &  \multicolumn{2}{c}{ Dead (Other) } &  \multicolumn{2}{c}{ Dead (BC)}  \\ 
  \hline
    &  \multicolumn{2}{c}{State 4} &  \multicolumn{2}{c}{State 5}  &  \multicolumn{2}{c}{State 4} &  \multicolumn{2}{c}{State 5}  \\ 
   \hline
  & M & S-M & M & S-M & M & S-M & M & S-M \\ 
  \hline
   S1 &  & &  &  & & &  &  \\ 
65--69 & 363 & 363 &   0 &   0 & 7003 & 7010 &  -8 &   0 \\ 
70--74 & 608 & 607 &  $\mathit{-1}$ &  $\mathit{-1}$ & 9301 & 9293 & $\mathit{-11}$ & $\mathit{-15}$ \\ 
75--79 & 1012 & 1012 &  $\mathit{-1}$&  $\mathit{-2 }$& 11767 & 11770 & $\mathit{-16}$ & $\mathit{-23}$ \\ 
80--84 & 1700 & 1700 &  $\mathit{-3}$ &  $\mathit{-4}$ & 14350 & 14348 & $\mathit{-25}$ & $\mathit{-34}$ \\ 
 85--89 & 2255 & 2255 &  $\mathit{-5}$ & $ \mathit{-6}$ & 13167 & 13169 & $\mathit{-27}$ & $\mathit{-35}$ \\ 
   S2 &  & &  &  & & &  &  \\ 
 65--69 & 363 & 363 &   8 &  10 & 7000 & 7010 & 152 & 193 \\ 
 70--74 & 607 & 607 &   7 &   9 & 9298 & 9293 & 113 & 138 \\ 
75--79 & 1011 & 1012 &   8 &  10 & 11762 & 11770 &  92 & 116 \\ 
80--84 & 1699 & 1699 &   7 &   9 & 14342 & 14340 &  63 &  76 \\ 
85--89 & 2253 & 2253 &   5 &   6 & 13158 & 13158 &  29 &  35 \\ 
   \hline
\end{tabular}}
{\parbox{5.3in}{
		\footnotesize Note: Results are based on Markov (M) and semi-Markov (S-M) models in the pandemic scenarios, Scenario 1 (S1) and Scenario 2 (S2), as compared to the pre-pandemic calibration, for $\alpha=0.6$, $\mu^{13} = \frac{1}{7}\mu^{23}$.
}
}
\end{table}

\subsection{{Years of life lost}}
\mbox{} \\
We calculate age-specific years of life lost (YLL) from BC and other causes at a given time $t$, denoted by $\text{YLL}^{\text{cause}}_{x,t}$, as 

\begin{eqnarray}
	\text{YLL}^{\text{cause}}_{x,t} = D^{\text{cause}}_{x,t} e_x,
	\label{eq:YLL1}
\end{eqnarray} 

\noindent where $D^{\text{cause}}_{x,t}$ shows the corresponding excess deaths from a given cause, and
$e_x$ is a function that quantifies the number of years lost for deceased people aged $x$ at time of death.
Here $e_x$ is determined as average life expectancy at age $x$ using standard life tables \citep{WHO2013}.
Also, total YLL for all ages, $\text{YLL}^{\text{cause}}_{t}$, are calculated as  

\begin{eqnarray}
	\text{YLL}^{\text{cause}}_{t} = \sum_{x} { D^{\text{cause}}_{x,t} e_x }.
	\label{eq:YLL2}
\end{eqnarray}
We refer to standard life tables as a source for the years loss function, following \citet{WHO2013}. 
Particularly, we use the 2018--2020 national standard life tables for women in the UK, with the life expectancies for women for ages 65--89, $e_x$,  shown in \tabref{tab:LEONSEstimates} \citep{ONSLifeExpTable}.

\begin{table}[H]
	\centering 
	\caption{
		{ Average life expectancies at various ages, denoted by $e_x$. \label{tab:LEONSEstimates}}}
\stackunder{	\begin{tabular}{cccccc}
		\hline
		Age &65--69 & 70--74 & 75--79 & 80--84 & 85--89 \\ \hline
		$e_x$&19.31 & 15.31 & 11.63& 8.44 & 5.84 \\ \hline
	\end{tabular}}
{\parbox{3in}{
		\footnotesize Note: The values are based on the 2018--2020 national standard life tables.\\
		Source: See \citet{ONSLifeExpTable} for women.
	}
}
\end{table}

\tabref{tab:DifferencesbyAge} shows that the semi-Markov model gives more years of life lost due to BC, as compared to the Markov model. This is a direct result of the former model estimating higher numbers of death due to BC. 
For deaths from other causes, we found 7,000--14,350 years of life lost across Scenarios 1 and 2 under the Markov model, with almost identical results under the semi-Markov model (\tabref{tab:DifferencesbyAge}).


\section{Sensitivity Analysis} \label{Sec:SensitivityAnalysis}

In this section we assess the sensitivity of our main findings to the values of certain model parameters.
{\tabref{tab:ParameterValues} shows different parametrisation for pre-pandemic model calibration and pandemic scenarios.}

\begin{table}[H]
\centering
\caption{{Parameter values applied in different sections}.}
\begin{small}
	\begin{tabular}{ll}
		\toprule
		Parametrisation in Section~\ref{secc:Results}& $\alpha=0.6$, $\beta=1/7$\\
		Parametrisation in Section~\ref{Sec:SensitivityAnalysis}& $\alpha=0.4$, $\beta=1/7$ \\
		& $\alpha=0.8$, $\beta=1/7$ \\
		& $\alpha=0.6$, $\beta=1/5$ \\
		& $\alpha=0.6$, $\beta=1/10$ \\
		& $\alpha=0.6$, $\beta=1/7$, \text{20\% lower value for $\mu^{35}_x$}  \\
		& $\alpha=0.6$, $\beta=1/7$, \text{20\% higher value for $\mu^{35}_x$}  \\
		\bottomrule
	\end{tabular}%
\end{small}
\label{tab:ParameterValues}%
\end{table} 

\subsection{{Impact of  parameter $\alpha$}}\label{sec:Sensitivityalpha}
\mbox{} \\

{In the pre-pandemic calibration and the pandemic scenarios in Section~\ref{secc:Results}}, it was assumed that 60\% of women developing BC, would actually be diagnosed with BC, in a given year, by choosing $\alpha = 0.6$ (Section~\ref{secc:Results}). 
{We now vary the value of $\alpha$, while keeping all other model characteristics fixed in the pre-pandemic calibration and pandemic scenarios.}
Higher and lower diagnosis rates are represented by assuming $\alpha = 0.8$ and $\alpha = 0.4$, respectively. 
Changing $\alpha$ mainly affects transitions to State 2 and State 3, along with smaller impacts on State 0 and State 5. 
For a woman free of BC, the probabilities of being in States 2--3 over 5 years have changed considerably {as compared to the pre-pandemic calibration when $\alpha = 0.6$ (\tabref{tab:TransitionsbyAge1year}).
Specifically, we observe an increase, mostly around 2 times higher}, when $\alpha=0.4$ and a decline, {by 70\% in ${_{5}} p^{02}_x$ and 50\% in ${_{5}} p^{03}_x$}, when $\alpha=0.8$. {Changes in State 0 and State 5} are more evident in the presence of lower diagnosis under both modelling settings.

Changes in excess deaths and YLL from other causes remain similar to those obtained for $\alpha=0.6$ (\tabref{tab:DifferencesbyAge}, \tabref{tab:DifferencesbyAge_Case1}--\tabref{tab:DifferencesbyAge_Case1b}). 
Considering excess deaths from BC, a lower pre-pandemic diagnosis rate of $\alpha=0.4$ leads to an increase of about 2\%, {corresponding 7 or less deaths across different ages, as compared to the corresponding pre-pandemic calibration}, in the Markov model, whereas the semi-Markov model suggests a slightly higher increase, about 3\%, {around and less than 10 deaths at the same ages}.
Meanwhile, a higher diagnosis rate of $\alpha=0.8$ leads to a more dramatic increase in BC deaths, which is about 9--12\%, {corresponding 7--11 excess deaths}, for the same ages under both models.

\subsection{{Impact of parameter $\beta$}}\label{sec:Sensitivitybeta}
\mbox{} \\
{In the pre-pandemic calibration and the pandemic scenarios in Section~\ref{secc:Results}}, we assumed $\beta$ as low as $\frac{1}{7}$, assuming that the transition from State 2 to State 3, $\mu^{23}_{x, z}$, can be 7 times higher than the transition from State 1 to State 3, $\mu^{13}_{x, z}$. This is mainly motivated by the absence of treatment in State 2, along with the potential pace of BC tumour growth (Section~\ref{sec:TumourGrowthBC}). 
All else being equal, we vary the value of $\beta$ by replacing it with $\frac{1}{5}$ and $\frac{1}{10}$.
{Note that there is no change in $\mu^{13}_{x, z}$, with $\mu^{13}_{x} = 0.01954$ in the Markov model, or determined by \eqref{eq:PolyFuncMu13} in the semi-Markov model (Section~\ref{sec:Determiningmu13}).}
Similar to Section~\ref{sec:Sensitivityalpha}, the main impact of changes in $\beta$ appears to be on State 2 and State 3, with higher changes occurring when $\beta = \frac{1}{10}$.
A smaller value of $\beta$ leads to more transitions into State 3, leaving a smaller number of women in State 2 {in the relevant pre-pandemic model calibration} (\tabref{tab:TransitionsbyAge1year}). 
The numbers in State 5 increase with a decreasing level of $\beta$ over time, because of the higher numbers of women with advanced BC (Stage 4 BC) in State 3.

\tabref{tab:DifferencesbyAge_Case2} and \tabref{tab:DifferencesbyAge_Case2b} show comparable outcomes for excess deaths and YLL from other causes. 
Excess deaths, along with YLL, from BC differ slightly from those obtained when $\beta= \frac{1}{7}$. 
{For a relatively higher value of $\beta $, $\frac{1}{5}$, BC deaths are around 2--5\% higher across different ages, 
indicating 3--7 excess deaths, as compared to the corresponding pre-pandemic calibrations, in both modelling settings. For a smaller value of $\beta$, $\frac{1}{10}$, deaths are around 3--6\% higher than the relevant pre-pandemic calibrations, corresponding to 7--14 excess deaths at different ages.}

\subsection{{Impact of transitions to death from breast cancer $\mu^{35}_x$}}\label{sec:Sensitivitykappa}
\mbox{} \\
{In the pre-pandemic calibration and the pandemic scenarios in Section~\ref{secc:Results}}, we assumed the transition to death from BC, $\mu^{35}_x$, to follow the rates reported in \tabref{tab:TransitionIntensities1.v4}. 
We now consider $\mu^{35}_x$ to be 20\% lower, or 20\% higher, than the rates in  \tabref{tab:TransitionIntensities1.v4}, {where the pre-pandemic model calibrations in these cases are shown in \tabref{tab:TransitionsbyAge1year}.}
The main effect of a change in this particular transition intensity is on cancer mortality (State 5), and on State 3.
For instance, an increase in the level of $\mu^{35}_x$ leads to a decrease in the number of women in State 3 and an increase in State 5. 
A considerable increase {in 5-year cancer mortality, ${_{5}} p^{15}_x$ and ${_{5}} p^{35}_x$, corresponding less than 11\% and 8\% increase across different ages, respectively,}
is also observed as a result of a higher level of $\mu^{35}_x$. 
This leads to a higher level of overall mortality, as well. 
The changes in cancer mortality are more evident for women with advanced BC. 

Similarly to Sections~\ref{sec:Sensitivityalpha}--\ref{sec:Sensitivitybeta}, varying rates of $\mu^{35}_x$
mainly results in changes in the number of excess BC deaths, while other outcomes, e.g. excess deaths from other causes, have remained comparable to the ones in Section~\ref{secc:Results}. 
An increasing level of $\mu^{35}_x$ leads to higher
number of excess BC deaths, {5--12 deaths across different ages}, 
whereas a decreasing level of $\mu^{35}_x$ results in smaller number of BC deaths, 
{4--9 excess deaths at the same ages}, with a similar effect on YLL from BC. 
However, the relative increase across ages, in comparison to the relevant pre-pandemic calibration, has remained the same, 3--6\%, independent of the level of $\mu^{35}_x$, under both models (\tabref{tab:DifferencesbyAge_Case3}, \tabref{tab:DifferencesbyAge_Case3b}).

We also obtain cancer survival probabilities, up to 10 years, for different values of $\mu^{35}_x$, provided in Appendix~\ref{app:CancerSurvival_Case3}. 
Note that different values of $\alpha$ and $\beta$ are not relevant to this calculation.
Consistent with the findings in \tabref{tab:CancerSpecificSurvRatesbyAge}, 
\tabref{tab:CancerSpecificSurvRatesbyAge_Case3a} and \tabref{tab:CancerSpecificSurvRatesbyAge_Case3b} 
point towards higher changes in cancer survival for women with pre-metastatic BC using different modelling settings, 
with these changes becoming more profound in time. 
Although an increasing level of $\mu^{35}_x$ results in lower cancer survival for women with metastatic BC, 
our model still suggests smaller differences between cancer survival at the oldest and youngest age groups in comparison to ONS methodology.


\section{Discussion} \label{sec:Conclusion}

During national lockdowns, essential BC diagnostic services were severely affected, along with cancer referral pathways. 
Health-seeking behaviour was also adversely affected, as only patients with urgent concerns were encouraged to use available services \citep{Maringeetal2020}. 
It is therefore important to further examine possible implications of late diagnoses on cancer rates and excess deaths. 

We have constructed a semi-Markov model to quantify changes in BC mortality for women aged 65+, 
as a result of the impact of COVID-19 on health services. 
\citet{Maringeetal2020} noted a 7.9--9.6\% increase in the number of deaths due to BC in a 5-year period after diagnosis, assuming that cancers could only be diagnosed through urgent referrals with up to 80\% reductions in cancer referrals. 
We assume 20\% reduction in BC diagnosis based on a more recently published report \citep{PHS2021}. 
As a result, we found a 3--6\% increase in the number of deaths from BC at different ages, and a 5--8\% increase in deaths from other causes, {as compared to the pre-pandemic model calibration in Section~\ref{secc:Results}. 
Also, our results showed considerable differences among certain occupancy probabilities, e.g. ${_{5}} p^{15}_x$, between the semi-Markov and Markov models, highlighting the significance of assuming duration dependence in the modelling.}

\subsection{{Strengths and limitations}}
\mbox{} \\
Low availability of suitable data was a major challenge in this study, limiting 
our ability to make data-driven inferences and to quantify uncertainty through appropriate statistical measures. 
A related key issue was the incompleteness of BC stage information in population-based cancer data. 
Nevertheless, our models {are based on a pragmatic combination of available data, literature information and modelling assumptions.
The models have produced} insightful findings, while the results are broadly consistent with existing literature. 
Our modelling approach has also provided estimates of excess deaths both from BC and from other causes.
Furthermore, sensitivity testing has been carried out to take into account parameter uncertainty to a certain extent. 
As expected, model outputs are sensitive to the choice of key model parameters. {Importantly, sensitivity to parameter $\alpha$ demonstrates the model's ability to capture the impact of health-service disruptions to BC mortality.} 
{Relative changes in cancer mortality and deaths from other causes have shown consistent results based on different parametrisations in various pre-pandemic model calibrations and pandemic scenarios. }

Our approach provides a valuable model, relating to delays in the provision of BC diagnostic and treatment services, which can be more accurately calibrated as more data become available. Availability of more data can help expand the modelling setting by providing more information in relation to the progression of BC.
Our model can also be used to represent different levels of BC service availability in non-pandemic times and therefore also provides a framework for comparing health service provision in different countries. 
It can allow further insights regarding the impact of a pandemic on different health services 
by changing the levels of $\alpha$ and $\beta$ parameters.

There are important areas for further research. The modelling framework can be extended in a number of ways, including the following:

\begin{itemize}
	\item  employing a more detailed clinical model for BC, e.g. by involving locally advanced BC and/or considering treatment and recovery options, which would allow distinguishing between recurrence of non-metastatic BC and developing of metastatic BC;
	\item considering multi-morbidity as an underlying condition, allowing for the potential impact on excess deaths;
	\item introducing time trend for BC mortality and morbidity over years;
	\item formally measuring parameter and model uncertainty.
\end{itemize}

\subsection{{Implications of this research}}
\mbox{} \\
Our study can inform decision makers by increasing awareness about the continuing impact of the COVID-19 pandemic. The estimated results can be helpful while implementing evidence-based health interventions. 

Our findings can also help life insurers understand the impact of late diagnoses or prevented treatment of a major cancer in women, on cancer mortality and survival rates. 
The modelling framework developed here can be useful for assessing different scenarios of cancer diagnoses, not just under pandemic circumstances, but also given different levels of health service provision. 

Our work can add value while considering insurance pricing and valuation assumptions {\color{black} related to ages of 65 years and over, which is also important for pension plans and healthcare at advanced age.}
{\color{black} Our model can be particularly relevant to critical illness and life insurance. For example, BC is one of the most common causes of critical illness claims among women, accounting for 44\% of all claims in 2014 in the UK \citep{CMI2011, Aviva2015}. 
In addition, our approach provides a more detailed modelling framework, as compared to one currently used by the insurance industry (\citet{ReynoldsFaye2016}), and can therefore provide better insights in relation to insurance cash flows. 
For instance, \citet{Ariketal2023b} show that accounting for duration-dependent rates in BC progression can have an impact on actuarial net premiums, affecting short and long-term insurance products differently, while a semi-Markov model leads to intuitive results aligned with the medical literature. 
}

Increases in population longevity, together with the relatively and increasingly long BC survival, mean that BC will continue to significantly affect older women \citep{Shacharetal2016, BCRF2021}. 
In this article we have explored the short-term impact of COVID-19 related BC diagnostic delays on related mortality in an older population. 

\section*{Acknowledgements}

ED and GS acknowledge funding from the Society of Actuaries, under a research project entitled `Predictive Modelling for Medical Morbidity Trends related to Insurance'. 
AA and GS acknowledge funding from SCOR Foundation for Science, under a project entitled `Estimating The Impact Of The COVID-19 Pandemic On Breast Cancer Deaths - An Application On Breast Cancer Life Insurance'.

\clearpage
\Urlmuskip=0mu plus 1mu\relax
\bibliographystyle{abbrvnat}
\bibliography{reference}

\paragraph{\textbf{Corresponding author}}
\mbox{} \\
Ay\c{s}e Ar{\i}k

E-mail: {A.ARIK@hw.ac.uk}

Address: {Department of Actuarial Mathematics and Statistics, 
	Heriot-Watt University, and the Maxwell Institute for Mathematical Sciences, 
	UK.}

\paragraph{\textbf{Competing interests}}
\mbox{} \\
The author(s) declare none.

\clearpage
\begin{appendices}
	\renewcommand\thetable{\thesection\arabic{table}}
	\renewcommand\thefigure{\thesection\arabic{figure}}
	
	\section{Modified Kolmogorov equations with duration dependence for breast cancer model}\label{sec:AppendixKolmogorov}

Modified Kolmogorov equations for the 6-state BC model, in \figref{fig:BC_model}, are given as below. 
Note that more details can be found in \citet{CMI1991}, based on a 3-state multiple model, allowing recovery from the disease under inspection along with duration dependence.
Here, in order to make integrals clearer, we introduce actuarial selection notation. 
For instance, $\mu^{13}_{x,z}$ is shown based on select attained age $[x]$ with duration $z$, specifically $\mu^{13}_{x,z} = \mu^{13}_{[x]+z}$.

\begin{subequations}
	\label{eq:KolmogorovSemi_markov}
	\begin{align}
		\frac{d}{dt}	{_ {t}} p^{00}_x  &= - {_ t} p^{00}_x \,    \bigg(   \mu^{01}_{x+t}   + \mu^{02}_{x+t} +\mu^{04}_{x+t}   \bigg)  \\
		\frac{d}{dt}	{_ {t}} p^{01}_{x}  &= {_ {t}} p^{00}_x \mu^{01}_{x+t} - {_ t} p^{01}_{x} \, \,  \mu^{14}_{x+t} 
		- \int_{u=0}^t{  {_ {u}} p^{00}_x \,\, \mu^{01}_{x+u} \,\,  {_ {t-u}} p^{11}_{[x+u]} \,\,  \mu^{13}_{[x+u]+t-u}\,\, du  }  \\
		\frac{d}{dt}	{_ {t}} p^{02}_x  &=  {_ {t}} p^{00}_x \mu^{02}_{x+t}  - {_ {t}} p^{02}_x \mu^{24}_{x+t}   - 
		\int_{u=0}^{t} {   {_ {u}} p^{00}_x \,\, \mu^{02}_{x+u} \,\,  {_ {t-u}} p^{22}_{[x+u]} \,\,  \mu^{23}_{[x+u]+t-u}\,\, du  } \\
		\frac{d}{dt}	{_ {t}} p^{03}_x  &=  	\int_{u=0}^{t} {   {_ {u}} p^{00}_x \,\, \mu^{01}_{x+u} \,\,  {_ {t-u}} p^{11}_{[x+u]} \,\,  \mu^{13}_{[x+u]+t-u}\,\, du  } +\\\nonumber
		& \int_{u=0}^{t} {   {_ {u}} p^{00}_x \,\, \mu^{02}_{x+u} \,\,  {_ {t-u}} p^{22}_{[x+u]} \,\,  \mu^{23}_{[x+u]+t-u}\,\, du  } 
		- {_ t} p^{03}_x \, \bigg(    \mu^{34}_{x+t} + \mu^{35}_{x+t}  \bigg) \\
		\frac{d}{dt}	{_ {t}} p^{04}_x  &= {_ {t}} p^{00}_x \mu^{04}_{x+t} +  {_ {t}} p^{01}_x \mu^{14}_{x+t}  +
		{_ {t}} p^{02}_x \mu^{24}_{x+t}  + {_ {t}} p^{03}_x \mu^{34}_{x+t}  \\
		\frac{d}{dt}	{_ {t}} p^{05}_x  &=  {_ {t}} p^{03}_x \mu^{35}_{x+t} 
	\end{align}
\end{subequations}

\newpage
{We note that the select notation on age $[x]$ is kept in the equations below, where this is based on the assumption of being in the relevant initial state. }

\begin{subequations}
	\label{eq:Kolmogorov_markov}
	\begin{align}
		\frac{d}{dt}	{_ {t}} p^{11}_{[x]}  &= - {_ t} p^{11}_{[x]} \, \bigg(  \mu^{13}_{[x]+t}   +  \mu^{14}_{[x]+t} \bigg)  \\
		\frac{d}{dt}	{_ {t}} p^{13}_{[x] } &=   {_ {t}} p^{11}_{[x]} \,\, \mu^{13}_{[x]+t}   
		-    {_ {t}} p^{13}_{[x]} \mu^{34}_{[x]+t} 
		-    {_ {t}} p^{13}_{[x]} \mu^{35}_{[x]+t}  \\
		\frac{d}{dt}	{_ {t}} p^{14}_{[x] } &= {_ {t}} p^{11}_{[x]}\,\, \mu^{14}_{[x]+t}  
		+  {_ {t}} p^{13}_{[x]} \mu^{34}_{[x]+t}   \\ 
		\frac{d}{dt}	{_ {t}} p^{15}_{[x]}  &=  	 {_ {t}} p^{13}_{[x]} \mu^{35}_{[x]+t} \\
			\frac{d}{dt}	{_ {t}} p^{22}_{[x]}  &= - {_ t} p^{22}_{[x]} \, \bigg(  \mu^{23}_{[x]+t}   +  \mu^{24}_{[x]+t} \bigg)  \\
		\frac{d}{dt}	{_ {t}} p^{23}_{[x]}  &=  {_ {t}} p^{22}_{[x]} \,\, \mu^{23}_{[x]+t} - {_ {t}} p^{23}_{[x]}\,  \bigg(   \mu^{34}_{[x]+t} +   \mu^{35}_{[x]+t}  \bigg) \\ 
			\frac{d}{dt}	{_ {t}} p^{24}_{[x]}  &=   {_ {t}} p^{22}_{[x]} \,\, \mu^{24}_{[x]+t} +  {_ {t}} p^{23}_{[x]} \,\, \mu^{34}_{[x]+t}  \\
		\frac{d}{dt}	{_ {t}} p^{25}_{[x]}  &=   {_ {t}} p^{23}_{[x]} \,\, \mu^{35}_{[x]+t} \\
		\frac{d}{dt}	{_ {t}} p^{33}_{[x]}  &= - {_ t} p^{33}_{[x]} \, \bigg(  \mu^{34}_{[x]+t}   +  \mu^{35}_{[x]+t} \bigg)  \\
			\frac{d}{dt}	{_ {t}} p^{34}_{[x]}  &=  {_ t} p^{33}_{[x]} \,\, \mu^{34}_{[x]+t}   \\
			\frac{d}{dt}	{_ {t}} p^{35}_{[x]}  &=  {_ t} p^{33}_{[x]} \,\, \mu^{35}_{[x]+t}  .
	\end{align} 
\end{subequations}

\afterpage{
\newgeometry{top=30mm, left=5mm, right=10mm, bottom=30mm,headsep=10mm, footskip=12mm}
\begin{landscape}
	
\section{Occupancy probabilities over 5 years in pre-pandemic model calibration}  \label{app:TransNumPrePandemic}

\begin{table}[H]
	\centering
\caption{Occupancy probabilities (\%) for women being in different states at the end of 5 years given that they have no breast cancer or clinically diagnosed with breast cancer at time zero based on Markov (M) and semi-Markov (S-M) models in the pre-pandemic model calibration using different choices of $\alpha$, $\beta$ parameters and $\mu^{35}$. 	\label{tab:TransitionsbyAge1year}}
	\resizebox{1\columnwidth}{!}{
\begin{tabular}{lrr>{\em}rr>{\em}rr>{\em}rrr>{\em}rr>{\em}rr>{\em}rrrr>{\em}rr>{\em}rrr}
  \hline
  & \multicolumn{10}{c}{From State 0} & \multicolumn{4}{c}{From State 1} & \multicolumn{2}{c}{From State 3}  & \multicolumn{4}{c}{From State 1} & \multicolumn{2}{c}{From State 3} \\
  \hline
  Age & $  {_{5}} p^{00}_x $& \multicolumn{2}{c}{ ${_{5}} p^{01}_x $}  &\multicolumn{2}{c}{ ${_{5}} p^{02}_x $} & \multicolumn{2}{c}{ ${_{5}} p^{03}_x $}& $ {_{5}} p^{04}_x $ &\multicolumn{2}{c}{ ${_{5}} p^{05}_x  $}  & \multicolumn{2}{c}{ ${_{1}} p^{15}_x  $} & \multicolumn{2}{c}{ ${_{5}} p^{15}_x  $} & $ {_{1}} p^{35}_x $ & $ {_{5}} p^{35}_x $ &  \multicolumn{2}{c}{ $ {_{1}} p^{14}_x  + {_{1}} p^{15}_x  $}  & \multicolumn{2}{c}{ $ {_{5}} p^{14}_x  + {_{5}} p^{15}_x  $} &  \multicolumn{2}{c} {$ {_{t}} p^{34}_x  + {_{t}} p^{35}_x  $ } \\ 
    \hline
  &\multicolumn{1}{c} {M}  & \multicolumn{1}{c} {M} & \multicolumn{1}{c} {S-M} & \multicolumn{1}{c} {M} & \multicolumn{1}{c} {S-M} & \multicolumn{1}{c} {M} & \multicolumn{1}{c} {S-M} & \multicolumn{1}{c} {M} & \multicolumn{1}{c} {M} & \multicolumn{1}{c} {S-M} & \multicolumn{1}{c} {M} & \multicolumn{1}{c} {S-M}&
  \multicolumn{1}{c} {M} & \multicolumn{1}{c} {S-M} & \multicolumn{1}{c} {M}& \multicolumn{1}{c} {M} &\multicolumn{1}{c} {M} & \multicolumn{1}{c} {S-M}  & \multicolumn{1}{c} {M} & \multicolumn{1}{c} {S-M} &\multicolumn{1}{c} {M, $t=1$} & \multicolumn{1}{c} {M, $t=5$}\\
  \hline
  	&	 \multicolumn{22}{c}{$\alpha=0.6$; $\mu^{13} = \frac{1}{7}\mu^{23}$ }  \\
  	\hline
65--69 & 93.09 & 1.50 & 1.47 & 0.76 & 0.68 & 0.24 & 0.31 & 4.29 & 0.13 & 0.16 & 0.25 & 0.16 & 4.24 & 5.98 & 24.36 & 74.15 & 1.12 & 1.03 & 8.47 & 10.18 & 25.12 & 76.47 \\ 
70--74 & 90.49 & 1.25 & 1.22 & 0.63 & 0.57 & 0.18 & 0.23 & 7.32 & 0.13 & 0.16 & 0.31 & 0.20 & 4.82 & 6.82 & 30.02 & 81.25 & 1.82 & 1.71 & 12.00 & 13.96 & 31.29 & 84.68 \\ 
75--79 & 85.07 & 1.33 & 1.31 & 0.67 & 0.61 & 0.18 & 0.24 & 12.59 & 0.15 & 0.19 & 0.34 & 0.22 & 4.92 & 6.97 & 32.56 & 82.61 & 2.99 & 2.88 & 17.27 & 19.24 & 34.75 & 88.17 \\ 
80--84 & 75.07 & 1.29 & 1.26 & 0.65 & 0.59 & 0.15 & 0.20 & 22.66 & 0.17 & 0.21 & 0.40 & 0.26 & 5.09 & 7.21 & 38.26 & 84.79 & 5.40 & 5.27 & 27.26 & 29.22 & 42.22 & 93.56 \\ 
85--89 & 59.71 & 1.09 & 1.07 & 0.55 & 0.50 & 0.13 & 0.17 & 38.36 & 0.16 & 0.19 & 0.39 & 0.25 & 4.47 & 6.29 & 37.65 & 79.54 & 9.61 & 9.47 & 42.05 & 43.62 & 44.94 & 94.94 \\ 
\hline
  	&	 \multicolumn{22}{c}{$\alpha=0.8$; $\mu^{13} = \frac{1}{7}\mu^{23}$ }  \\
\hline
 65--69 & 93.73 & 1.50 & 1.47 & 0.29 & 0.26 & 0.12 & 0.16 & 4.29 & 0.07 & 0.08 & 0.25 & 0.16 & 4.24 & 5.98 & 24.36 & 74.15 & 1.12 & 1.03 & 8.47 & 10.18 & 25.12 & 76.47 \\ 
 70--74 & 91.04 & 1.25 & 1.23 & 0.24 & 0.21 & 0.09 & 0.12 & 7.32 & 0.06 & 0.08 & 0.31 & 0.20 & 4.82 & 6.82 & 30.02 & 81.25 & 1.82 & 1.71 & 12.00 & 13.96 & 31.29 & 84.68 \\ 
 75--79 & 85.65 & 1.34 & 1.31 & 0.25 & 0.23 & 0.09 & 0.12 & 12.60 & 0.08 & 0.09 & 0.34 & 0.22 & 4.92 & 6.97 & 32.56 & 82.61 & 2.99 & 2.88 & 17.27 & 19.24 & 34.75 & 88.17 \\ 
 80--84 & 75.63 & 1.29 & 1.27 & 0.25 & 0.22 & 0.08 & 0.11 & 22.66 & 0.09 & 0.11 & 0.40 & 0.26 & 5.09 & 7.21 & 38.26 & 84.79 & 5.40 & 5.27 & 27.26 & 29.22 & 42.22 & 93.56 \\ 
 85--89 & 60.18 & 1.09 & 1.07 & 0.21 & 0.19 & 0.07 & 0.09 & 38.37 & 0.08 & 0.10 & 0.39 & 0.25 & 4.47 & 6.29 & 37.65 & 79.54 & 9.61 & 9.47 & 42.05 & 43.62 & 44.94 & 94.94 \\  
 \hline
 &	 \multicolumn{22}{c}{$\alpha=0.4$; $\mu^{13} = \frac{1}{7}\mu^{23}$ }  \\
 \hline
65--69 & 91.80 & 1.49 & 1.46 & 1.69 & 1.52 & 0.47 & 0.61 & 4.29 & 0.26 & 0.32 & 0.25 & 0.16 & 4.24 & 5.98 & 24.36 & 74.15 & 1.12 & 1.03 & 8.47 & 10.18 & 25.12 & 76.47 \\ 
 70--74 & 89.42 & 1.24 & 1.22 & 1.41 & 1.27 & 0.35 & 0.46 & 7.32 & 0.26 & 0.32 & 0.31 & 0.20 & 4.82 & 6.82 & 30.02 & 81.25 & 1.82 & 1.71 & 12.00 & 13.96 & 31.29 & 84.68 \\ 
 75--79 & 83.93 & 1.32 & 1.30 & 1.50 & 1.35 & 0.35 & 0.46 & 12.59 & 0.30 & 0.37 & 0.34 & 0.22 & 4.92 & 6.97 & 32.56 & 82.61 & 2.99 & 2.88 & 17.27 & 19.24 & 34.75 & 88.17 \\ 
 80--84 & 73.97 & 1.28 & 1.25 & 1.46 & 1.31 & 0.30 & 0.40 & 22.65 & 0.34 & 0.42 & 0.40 & 0.26 & 5.09 & 7.21 & 38.26 & 84.79 & 5.40 & 5.27 & 27.26 & 29.22 & 42.22 & 93.56 \\ 
85--89 & 58.78 & 1.08 & 1.06 & 1.23 & 1.10 & 0.26 & 0.34 & 38.34 & 0.31 & 0.38 & 0.39 & 0.25 & 4.47 & 6.29 & 37.65 & 79.54 & 9.61 & 9.47 & 42.05 & 43.62 & 44.94 & 94.94 \\ 
\hline
&	 \multicolumn{22}{c}{$\alpha=0.6$; $\mu^{13} = \frac{1}{5}\mu^{23}$ }  \\
\hline
65--69 & 93.09 & 1.50 & 1.47 & 0.83 & 0.76 & 0.19 & 0.26 & 4.29 & 0.10 & 0.13 & 0.25 & 0.16 & 4.24 & 5.98 & 24.36 & 74.15 & 1.12 & 1.03 & 8.47 & 10.18 & 25.12 & 76.47 \\ 
 70--74 & 90.49 & 1.25 & 1.22 & 0.69 & 0.64 & 0.14 & 0.19 & 7.32 & 0.10 & 0.13 & 0.31 & 0.20 & 4.82 & 6.82 & 30.02 & 81.25 & 1.82 & 1.71 & 12.00 & 13.96 & 31.29 & 84.68 \\ 
75--79 & 85.07 & 1.33 & 1.31 & 0.74 & 0.68 & 0.15 & 0.20 & 12.59 & 0.12 & 0.15 & 0.34 & 0.22 & 4.92 & 6.97 & 32.56 & 82.61 & 2.99 & 2.88 & 17.27 & 19.24 & 34.75 & 88.17 \\ 
80--84 & 75.07 & 1.29 & 1.26 & 0.71 & 0.66 & 0.13 & 0.17 & 22.66 & 0.14 & 0.17 & 0.40 & 0.26 & 5.09 & 7.21 & 38.26 & 84.79 & 5.40 & 5.27 & 27.26 & 29.22 & 42.22 & 93.56 \\ 
85--89 & 59.71 & 1.09 & 1.07 & 0.60 & 0.55 & 0.11 & 0.14 & 38.36 & 0.12 & 0.16 & 0.39 & 0.25 & 4.47 & 6.29 & 37.65 & 79.54 & 9.61 & 9.47 & 42.05 & 43.62 & 44.94 & 94.94 \\ 
\hline
&	 \multicolumn{22}{c}{$\alpha=0.6$; $\mu^{13} = \frac{1}{10}\mu^{23}$ }  \\
\hline
65--69 & 93.09 & 1.50 & 1.47 & 0.67 & 0.58 & 0.29 & 0.37 & 4.29 & 0.17 & 0.20 & 0.25 & 0.16 & 4.24 & 5.98 & 24.36 & 74.15 & 1.12 & 1.03 & 8.47 & 10.18 & 25.12 & 76.47 \\ 
70--74 & 90.49 & 1.25 & 1.22 & 0.56 & 0.49 & 0.22 & 0.28 & 7.32 & 0.16 & 0.20 & 0.31 & 0.20 & 4.82 & 6.82 & 30.02 & 81.25 & 1.82 & 1.71 & 12.00 & 13.96 & 31.29 & 84.68 \\ 
 75--79 & 85.07 & 1.33 & 1.31 & 0.59 & 0.52 & 0.22 & 0.28 & 12.59 & 0.19 & 0.23 & 0.34 & 0.22 & 4.92 & 6.97 & 32.56 & 82.61 & 2.99 & 2.88 & 17.27 & 19.24 & 34.75 & 88.17 \\ 
80--84 & 75.07 & 1.29 & 1.26 & 0.57 & 0.50 & 0.19 & 0.24 & 22.66 & 0.22 & 0.26 & 0.40 & 0.26 & 5.09 & 7.21 & 38.26 & 84.79 & 5.40 & 5.27 & 27.26 & 29.22 & 42.22 & 93.56 \\ 
85--89 & 59.71 & 1.09 & 1.07 & 0.49 & 0.42 & 0.16 & 0.21 & 38.35 & 0.20 & 0.24 & 0.39 & 0.25 & 4.47 & 6.29 & 37.65 & 79.54 & 9.61 & 9.47 & 42.05 & 43.62 & 44.94 & 94.94 \\ 
  \hline
  &	 \multicolumn{22}{c}{$\alpha=0.6$; $\mu^{13} = \frac{1}{7}\mu^{23}$; $\mu^{35}$ is 20\% lower than the baseline level }  \\
  \hline
65--69 & 93.09 & 1.50 & 1.47 & 0.76 & 0.68 & 0.26 & 0.33 & 4.29 & 0.11 & 0.14 & 0.20 & 0.13 & 3.66 & 5.14 & 20.02 & 66.26 & 1.07 & 1.00 & 7.90 & 9.36 & 20.80 & 68.85 \\ 
70--74 & 90.49 & 1.25 & 1.22 & 0.63 & 0.57 & 0.19 & 0.26 & 7.32 & 0.11 & 0.14 & 0.25 & 0.16 & 4.23 & 5.96 & 24.84 & 74.13 & 1.76 & 1.67 & 11.43 & 13.13 & 26.15 & 78.04 \\ 
75--79 & 85.07 & 1.33 & 1.31 & 0.67 & 0.61 & 0.20 & 0.26 & 12.59 & 0.13 & 0.16 & 0.28 & 0.18 & 4.34 & 6.12 & 27.04 & 75.96 & 2.93 & 2.84 & 16.72 & 18.43 & 29.32 & 82.35 \\ 
80--84 & 75.07 & 1.29 & 1.26 & 0.65 & 0.59 & 0.17 & 0.23 & 22.66 & 0.15 & 0.19 & 0.33 & 0.21 & 4.56 & 6.43 & 32.04 & 79.18 & 5.34 & 5.22 & 26.79 & 28.53 & 36.18 & 89.42 \\ 
85--89 & 59.71 & 1.09 & 1.07 & 0.55 & 0.50 & 0.15 & 0.19 & 38.36 & 0.14 & 0.17 & 0.32 & 0.21 & 4.00 & 5.60 & 31.52 & 73.80 & 9.54 & 9.43 & 41.68 & 43.06 & 39.15 & 91.67 \\ 
 \hline
   &	 \multicolumn{22}{c}{$\alpha=0.6$; $\mu^{13} = \frac{1}{7}\mu^{23}$; $\mu^{35}$ is 20\% higher than the baseline level }  \\
   \hline 
 65--69 & 93.09 & 1.50 & 1.47 & 0.76 & 0.68 & 0.22 & 0.29 & 4.29 & 0.15 & 0.18 & 0.29 & 0.19 & 4.73 & 6.70 & 28.47 & 80.14 & 1.16 & 1.06 & 8.95 & 10.89 & 29.21 & 82.23 \\ 
70--74 & 90.49 & 1.25 & 1.22 & 0.63 & 0.57 & 0.16 & 0.21 & 7.32 & 0.15 & 0.18 & 0.36 & 0.24 & 5.31 & 7.53 & 34.83 & 86.28 & 1.87 & 1.75 & 12.48 & 14.65 & 36.06 & 89.32 \\ 
 75--79 & 85.07 & 1.33 & 1.31 & 0.67 & 0.61 & 0.16 & 0.22 & 12.59 & 0.17 & 0.21 & 0.39 & 0.26 & 5.39 & 7.65 & 37.65 & 87.18 & 3.04 & 2.92 & 17.71 & 19.88 & 39.76 & 92.07 \\ 
80--84 & 75.07 & 1.29 & 1.26 & 0.65 & 0.59 & 0.14 & 0.18 & 22.66 & 0.19 & 0.24 & 0.46 & 0.31 & 5.51 & 7.82 & 43.90 & 88.46 & 5.46 & 5.32 & 27.63 & 29.76 & 47.68 & 96.08 \\ 
85--89 & 59.71 & 1.09 & 1.07 & 0.55 & 0.50 & 0.12 & 0.15 & 38.36 & 0.17 & 0.21 & 0.45 & 0.30 & 4.85 & 6.83 & 43.21 & 83.46 & 9.66 & 9.52 & 42.34 & 44.04 & 50.18 & 96.93 \\ 
     \hline 
\end{tabular}
}
\end{table}

\begin{table}[H]
	\centering
	\caption{Occupancy probabilities (\%) for women being in different states at the end of 1 year given that they have no breast cancer or clinically diagnosed with breast cancer at time zero based on Markov (M) and semi-Markov (S-M) models when $\alpha=0.6$; $\mu^{13} = \frac{1}{7}\mu^{23}$. 	\label{tab:TransitionsbyAge1year_CovidScenarios}}
	\resizebox{1\columnwidth}{!}{
		\begin{tabular}{lrr>{\em}rr>{\em}rr>{\em}rrr>{\em}rr>{\em}rr>{\em}rrrr>{\em}rr>{\em}rrr}
			\hline
			& \multicolumn{10}{c}{From State 0} & \multicolumn{4}{c}{From State 1} & \multicolumn{2}{c}{From State 3}  & \multicolumn{4}{c}{From State 1} & \multicolumn{2}{c}{From State 3} \\
			\hline
			Age & $  {_{1}} p^{00}_x $& \multicolumn{2}{c}{ ${_{1}} p^{01}_x $}  &\multicolumn{2}{c}{ ${_{1}} p^{02}_x $} & \multicolumn{2}{c}{ ${_{1}} p^{03}_x $}& $ {_{1}} p^{04}_x $ &\multicolumn{2}{c}{ ${_{1}} p^{05}_x  $}  & \multicolumn{2}{c}{ ${_{1}} p^{15}_x  $} & \multicolumn{2}{c}{ ${_{5}} p^{15}_x  $} & $ {_{1}} p^{35}_x $ & $ {_{5}} p^{35}_x $ &  \multicolumn{2}{c}{ $ {_{1}} p^{14}_x  + {_{1}} p^{15}_x  $}  & \multicolumn{2}{c}{ $ {_{5}} p^{14}_x  + {_{5}} p^{15}_x  $} &  \multicolumn{2}{c} {$ {_{t}} p^{34}_x  + {_{t}} p^{35}_x  $ } \\ 
			\hline
			&\multicolumn{1}{c} {M}  & \multicolumn{1}{c} {M} & \multicolumn{1}{c} {S-M} & \multicolumn{1}{c} {M} & \multicolumn{1}{c} {S-M} & \multicolumn{1}{c} {M} & \multicolumn{1}{c} {S-M} & \multicolumn{1}{c} {M} & \multicolumn{1}{c} {M} & \multicolumn{1}{c} {S-M} & \multicolumn{1}{c} {M} & \multicolumn{1}{c} {S-M}&
			\multicolumn{1}{c} {M} & \multicolumn{1}{c} {S-M} & \multicolumn{1}{c} {M}& \multicolumn{1}{c} {M} &\multicolumn{1}{c} {M} & \multicolumn{1}{c} {S-M}  & \multicolumn{1}{c} {M} & \multicolumn{1}{c} {S-M} &\multicolumn{1}{c} {M, $t=1$} & \multicolumn{1}{c} {M, $t=5$}\\
			\hline
			&	 \multicolumn{22}{c}{Pre-pandemic calibration}  \\
			\hline
65--69 & 98.58 & 0.33 & 0.33 & 0.21 & 0.21 & 0.02 & 0.01 & 0.87 &   0 &   0 & 0.25 & 0.16 & 4.24 & 5.98 & 24.36 & 74.15 & 1.12 & 1.03 & 8.47 & 10.18 & 25.12 & 76.47 \\ 
70--74 & 98.02 & 0.28 & 0.28 & 0.18 & 0.18 & 0.01 & 0.01 & 1.51 &   0 &   0 & 0.31 & 0.20 & 4.82 & 6.82 & 30.02 & 81.25 & 1.82 & 1.71 & 12.00 & 13.96 & 31.29 & 84.68 \\ 
75--79 & 96.82 & 0.31 & 0.31 & 0.20 & 0.20 & 0.01 & 0.01 & 2.66 &   0 &   0 & 0.34 & 0.22 & 4.92 & 6.97 & 32.56 & 82.61 & 2.99 & 2.88 & 17.27 & 19.24 & 34.75 & 88.17 \\ 
80--84 & 94.43 & 0.33 & 0.33 & 0.21 & 0.21 & 0.02 & 0.01 & 5.01 &   0 &   0 & 0.40 & 0.26 & 5.09 & 7.21 & 38.26 & 84.79 & 5.40 & 5.27 & 27.26 & 29.22 & 42.22 & 93.56 \\ 
85--89 & 90.20 & 0.34 & 0.34 & 0.21 & 0.22 & 0.02 & 0.01 & 9.23 &   0 &   0 & 0.39 & 0.25 & 4.47 & 6.29 & 37.65 & 79.54 & 9.61 & 9.47 & 42.05 & 43.62 & 44.94 & 94.94 \\ 
			&	 \multicolumn{22}{c}{ Pandemic scenarios}  \\
			S1 &	 \multicolumn{21}{c}{}  \\
65--69 & 98.49 & 0.33 & 0.33 & 0.20 & 0.21 & 0.02 & 0.01 & 0.96 &   0 &   0 & 0.25 & 0.16 & 4.23 & 5.96 & 24.36 & 74.03 & 1.21 & 1.12 & 8.81 & 10.52 & 25.19 & 76.56 \\ 
70--74 & 97.88 & 0.28 & 0.28 & 0.17 & 0.18 & 0.01 & 0.01 & 1.66 &   0 &   0 & 0.31 & 0.20 & 4.80 & 6.79 & 30.00 & 81.03 & 1.96 & 1.85 & 12.58 & 14.53 & 31.39 & 84.78 \\ 
75--79 & 96.56 & 0.31 & 0.31 & 0.20 & 0.20 & 0.01 & 0.01 & 2.91 &   0 &   0 & 0.33 & 0.22 & 4.88 & 6.91 & 32.53 & 82.24 & 3.24 & 3.13 & 18.22 & 20.17 & 34.92 & 88.31 \\ 
80--84 & 93.96 & 0.33 & 0.33 & 0.21 & 0.21 & 0.02 & 0.01 & 5.49 &   0 &   0 & 0.40 & 0.26 & 5.02 & 7.10 & 38.20 & 84.15 & 5.88 & 5.74 & 28.87 & 30.78 & 42.51 & 93.70 \\ 
85--89 & 89.42 & 0.33 & 0.34 & 0.21 & 0.22 & 0.02 & 0.01 & 10.02 &   0 &   0 & 0.39 & 0.25 & 4.36 & 6.12 & 37.55 & 78.56 & 10.39 & 10.26 & 44.17 & 45.68 & 45.43 & 95.12 \\ 
			S2 &	 \multicolumn{21}{c}{}  \\
65--69 & 98.49 & 0.28 & 0.28 & 0.25 & 0.26 & 0.02 & 0.01 & 0.96 &   0 &   0 & 0.25 & 0.16 & 4.23 & 5.96 & 24.36 & 74.03 & 1.21 & 1.12 & 8.81 & 10.52 & 25.19 & 76.56 \\ 
70--74 & 97.88 & 0.24 & 0.24 & 0.21 & 0.22 & 0.01 & 0.01 & 1.66 &   0 &   0 & 0.31 & 0.20 & 4.80 & 6.79 & 30.00 & 81.03 & 1.96 & 1.85 & 12.58 & 14.53 & 31.39 & 84.78 \\ 
75--79 & 96.56 & 0.26 & 0.26 & 0.24 & 0.25 & 0.02 & 0.01 & 2.91 &   0 &   0 & 0.33 & 0.22 & 4.88 & 6.91 & 32.53 & 82.24 & 3.24 & 3.13 & 18.22 & 20.17 & 34.92 & 88.31 \\ 
80--84 & 93.96 & 0.28 & 0.28 & 0.26 & 0.26 & 0.02 & 0.01 & 5.49 &   0 &   0 & 0.40 & 0.26 & 5.02 & 7.10 & 38.20 & 84.15 & 5.88 & 5.74 & 28.87 & 30.78 & 42.51 & 93.70 \\ 
85--89 & 89.42 & 0.28 & 0.29 & 0.26 & 0.27 & 0.02 & 0.01 & 10.02 &   0 &   0 & 0.39 & 0.25 & 4.36 & 6.12 & 37.55 & 78.56 & 10.39 & 10.26 & 44.17 & 45.68 & 45.43 & 95.12 \\ 
			\hline			
		\end{tabular}
	}
\end{table}

\end{landscape}
}

\newpage
\section{Excess deaths and years of life expectancy lost at different age groups in Section~\ref{sec:Sensitivityalpha}}  \label{app:YLLExcessDeaths_Case1}

\begin{table}[H]
	\centering
	\caption{
		{ Age-specific excess number of deaths and years of life expectancy lost (YLL), per 100,000 women, based on Markov (M) and semi-Markov (S-M) models in the pandemic scenarios, Scenario 1 (S1) and Scenario 2 (S2), as compared to the pre-pandemic calibration, for {$\alpha=0.8$, $\mu^{13} = \frac{1}{7}\mu^{23}$}. 
			\label{tab:DifferencesbyAge_Case1}} 
	}
	\begin{tabular}{lr>{\em}rr>{\em}rr>{\em}rr>{\em}r}
		\hline
		& \multicolumn{4}{c}{Excess deaths} &  \multicolumn{4}{c}{YLL} \\ \hline
		Age &  \multicolumn{2}{c}{ Dead (Other) } &  \multicolumn{2}{c}{ Dead (BC)}  &  \multicolumn{2}{c}{ Dead (Other) } &  \multicolumn{2}{c}{ Dead (BC)}  \\ 
		\hline
		&  \multicolumn{2}{c}{State 4} &  \multicolumn{2}{c}{State 5}  &  \multicolumn{2}{c}{State 4} &  \multicolumn{2}{c}{State 5}  \\ 
		\hline
		& M & S-M & M & S-M & M & S-M & M & S-M \\ 
		\hline
		S1 &  & &  &  & & &  &  \\ 
65--69 & 363 & 363 &   0 &   0 & 7004 & 7010 &  $\mathit{-4}$ &   0 \\ 
70--74 & 608 & 608 &   0 &  $\mathit{-1 }$& 9302 & 9308 &  $\mathit{-5 }$& $\mathit{-15}$ \\ 
75--79 & 1012 & 1012 &  $\mathit{-1}$ & $\mathit{ -1} $& 11769 & 11770 &  $\mathit{-8}$ & $\mathit{-12 }$\\ 
80--84 & 1701 & 1700 & $ \mathit{-2}$ &  $\mathit{-2}$ & 14352 & 14348 & $\mathit{-13} $&$ \mathit{-17}$ \\ 
85--89 & 2255 & 2255 & $ \mathit{-2}$ &  $\mathit{-3 }$& 13168 & 13169 & $\mathit{-13 }$& $\mathit{-18} $\\ 
  		S2 &  & &  &  & & &  &  \\ 
65--69 & 363 & 363 &   8 &  10 & 7001 & 7010 & 155 & 193 \\ 
70--74 & 607 & 608 &   8 &  10 & 9299 & 9308 & 118 & 153 \\ 
75--79 & 1011 & 1011 &   9 &  11 & 11764 & 11758 & 100 & 128 \\ 
80--84 & 1700 & 1699 &   9 &  11 & 14344 & 14340 &  76 &  93 \\ 
85--89 & 2253 & 2253 &   7 &   9 & 13159 & 13158 &  43 &  53 \\ 
		\hline
	\end{tabular}
\end{table}

\begin{table}[H]
	\centering
	\caption{
		{ Age-specific excess number of deaths and years of life expectancy lost (YLL), per 100,000 women, based on Markov (M) and semi-Markov (S-M) models in the pandemic scenarios, Scenario 1 (S1) and Scenario 2 (S2), as compared to the pre-pandemic calibration, for {$\alpha=0.4$, $\mu^{13} = \frac{1}{7}\mu^{23}$}. 
			\label{tab:DifferencesbyAge_Case1b}} 
	}
	\begin{tabular}{lr>{\em}rr>{\em}rr>{\em}rr>{\em}r}
		\hline
		& \multicolumn{4}{c}{Excess deaths} &  \multicolumn{4}{c}{YLL} \\ \hline
		Age &  \multicolumn{2}{c}{ Dead (Other) } &  \multicolumn{2}{c}{ Dead (BC)}  &  \multicolumn{2}{c}{ Dead (Other) } &  \multicolumn{2}{c}{ Dead (BC)}  \\ 
		\hline
		&  \multicolumn{2}{c}{State 4} &  \multicolumn{2}{c}{State 5}  &  \multicolumn{2}{c}{State 4} &  \multicolumn{2}{c}{State 5}  \\ 
		\hline
		& M & S-M & M & S-M & M & S-M & M & S-M \\ 
		\hline
		S1 &  & &  &  & & &  &  \\ 
65--69 & 363 & 363 & $ \mathit{-1}$ &  $\mathit{-1}$ & 7001 & 7010 &$ \mathit{-15}$ & $\mathit{-19}$ \\ 
70--74 & 607 & 607 &  $\mathit{-1}$ &  $\mathit{-2}$ & 9299 & 9293 & $\mathit{-21} $& $\mathit{-31 }$\\ 
75--79 & 1012 & 1011 &  $\mathit{-3}$ &  $\mathit{-3} $& 11764 & 11758 & $\mathit{-33}$ & $\mathit{-35} $\\ 
80--84 & 1700 & 1699 & $ \mathit{-6}$ &  $\mathit{-8 }$& 14346 & 14340 & $\mathit{-51}$ & $\mathit{-68}$ \\ 
85--89 & 2254 & 2255 & $ \mathit{-9}$& $\mathit{-11}$ & 13165 & 13169 & $\mathit{-53}$ & $\mathit{-64}$ \\ 
  		S2 &  & &  &  & & &  &  \\ 
65--69 & 362 & 363 &   7 &  10 & 6999 & 7010 & 143 & 193 \\ 
70--74 & 607 & 607 &   7 &   8 & 9295 & 9293 & 102 & 122 \\ 
75--79 & 1011 & 1011 &   6 &   8 & 11759 & 11758 &  75 &  93 \\ 
80--84 & 1699 & 1698 &   4 &   6 & 14337 & 14331 &  38 &  51 \\ 
85--89 & 2253 & 2253 &   0 &   1 & 13155 & 13158 &   3 &   6 \\ 
		\hline
	\end{tabular}
\end{table}

\section{Excess deaths and years of life expectancy lost at different age groups in Section~\ref{sec:Sensitivitybeta}}  \label{app:YLLExcessDeaths_Case2}

\begin{table}[H]
	\centering
	\caption{
		{ Age-specific excess number of deaths and years of life expectancy lost (YLL), per 100,000 women, based on Markov (M) and semi-Markov (S-M) models in the pandemic scenarios, Scenario 1 (S1) and Scenario 2 (S2), as compared to the pre-pandemic calibration, for {$\alpha=0.6$, $\mu^{13} = \frac{1}{5}\mu^{23}$}. 
			\label{tab:DifferencesbyAge_Case2}} 
	}
	\begin{tabular}{lr>{\em}rr>{\em}rr>{\em}rr>{\em}r}
		\hline
		& \multicolumn{4}{c}{Excess deaths} &  \multicolumn{4}{c}{YLL} \\ \hline
		Age &  \multicolumn{2}{c}{ Dead (Other) } &  \multicolumn{2}{c}{ Dead (BC)}  &  \multicolumn{2}{c}{ Dead (Other) } &  \multicolumn{2}{c}{ Dead (BC)}  \\ 
		\hline
		&  \multicolumn{2}{c}{State 4} &  \multicolumn{2}{c}{State 5}  &  \multicolumn{2}{c}{State 4} &  \multicolumn{2}{c}{State 5}  \\ 
		\hline
		& M & S-M & M & S-M & M & S-M & M & S-M \\ 
		\hline
		S1 &  & &  &  & & &  &  \\ 
65--69 & 363 & 362 &   0 &  $\mathit{-1}$ & 7003 & 6990 &  $\mathit{-6 }$&$ \mathit{-19}$ \\ 
70--74 & 608 & 608 &  $\mathit{-1} $&   0 & 9301 & 9308 &  $\mathit{-8 }$&   0 \\ 
75--79 & 1012 & 1011 &  $\mathit{-1 }$&  $\mathit{-2}$ & 11768 & 11758 &$ \mathit{-13}$ & $\mathit{-23} $\\ 
80--84 & 1700 & 1701 & $ \mathit{-2} $&  $\mathit{-3 }$& 14351 & 14356 & $\mathit{-20} $& $ \mathit{-25} $\\ 
85--89 & 2255 & 2255 &  $\mathit{-4} $& $ \mathit{-5 }$& 13168 & 13169 &$ \mathit{-21 }$& $\mathit{-29} $\\ 
  		S2 &  & &  &  & & &  &  \\ 
65--69 & 363 & 362 &   6 &   7 & 7001 & 6990 & 106 & 135 \\ 
70--74 & 607 & 608 &   5 &   7 & 9299 & 9308 &  79 & 107 \\ 
75--79 & 1012 & 1011 &   5 &   7 & 11764 & 11758 &  63 &  81 \\ 
80--84 & 1700 & 1700 &   5 &   7 & 14345 & 14348 &  42 &  59 \\ 
85--89 & 2254 & 2253 &   3 &   4 & 13161 & 13158 &  18 &  23 \\ 
   \hline		
	\end{tabular}
\end{table}

\begin{table}[H]
	\centering
	\caption{
		{ Age-specific excess number of deaths and years of life expectancy lost (YLL), per 100,000 women, based on Markov (M) and semi-Markov (S-M) models in the pandemic scenarios, Scenario 1 (S1) and Scenario 2 (S2), as compared to the pre-pandemic calibration, for {$\alpha=0.6$, $\mu^{13} = \frac{1}{10}\mu^{23}$}. 
			\label{tab:DifferencesbyAge_Case2b}} 
	}
	\begin{tabular}{lr>{\em}rr>{\em}rr>{\em}rr>{\em}r}
		\hline
		& \multicolumn{4}{c}{Excess deaths} &  \multicolumn{4}{c}{YLL} \\ \hline
		Age &  \multicolumn{2}{c}{ Dead (Other) } &  \multicolumn{2}{c}{ Dead (BC)}  &  \multicolumn{2}{c}{ Dead (Other) } &  \multicolumn{2}{c}{ Dead (BC)}  \\ 
		\hline
		&  \multicolumn{2}{c}{State 4} &  \multicolumn{2}{c}{State 5}  &  \multicolumn{2}{c}{State 4} &  \multicolumn{2}{c}{State 5}  \\ 
		\hline
		& M & S-M & M & S-M & M & S-M & M & S-M \\ 
		\hline
		S1 &  & &  &  & & &  &  \\ 
65--69 & 363 & 362 &  $\mathit{-1 }$& $ \mathit{-1}$ & 7002 & 6990 &$ \mathit{-10} $& $\mathit{-19}$ \\ 
70--74 & 607 & 608 &  $\mathit{-1} $&  $\mathit{-1 }$& 9300 & 9308 & $\mathit{-13 }$& $\mathit{-15}$ \\ 
75--79 & 1012 & 1011 &  $\mathit{-2 }$&  $\mathit{-2 }$& 11766 & 11758 & $\mathit{-21} $& $\mathit{-23 }$\\ 
80--84 & 1700 & 1700 &  $\mathit{-4}$& $ \mathit{-5}$ & 14349 & 14348 & $\mathit{-32} $& $\mathit{-42} $\\ 
85--89 & 2255 & 2255 & $\mathit{ -6} $&  $\mathit{-7 }$& 13167 & 13169 & $\mathit{-34} $& $\mathit{-41} $\\ 
  		S2 &  & &  &  & & &  &  \\ 
65--69 & 362 & 362 &  11 &  13 & 6999 & 6990 & 210 & 251 \\ 
70--74 & 607 & 607 &  10 &  13 & 9295 & 9293 & 157 & 199 \\ 
75--79 & 1011 & 1011 &  11 &  14 & 11759 & 11758 & 128 & 163 \\ 
80--84 & 1699 & 1698 &  11 &  13 & 14337 & 14331 &  90 & 110 \\ 
85--89 & 2252 & 2252 &   7 &   9 & 13153 & 13152 &  43 &  53 \\ 
		\hline
	\end{tabular}
\end{table}

\section{Excess deaths and years of life expectancy lost at different age groups in Section~\ref{sec:Sensitivitykappa}}  \label{app:YLLExcessDeaths_Case3}

\begin{table}[H]
	\centering
	\caption{
		{ Age-specific excess number of deaths and years of life expectancy lost (YLL), per 100,000 women, based on Markov (M) and semi-Markov (S-M) models in the pandemic scenarios, Scenario 1 (S1) and Scenario 2 (S2), as compared to the pre-pandemic calibration, for {$\alpha=0.6$, $\mu^{13} = \frac{1}{7}\mu^{23}$, $\mu^{35}=0.8 \mu^{35}$}. 
			\label{tab:DifferencesbyAge_Case3}} 
	}
	\begin{tabular}{lr>{\em}rr>{\em}rr>{\em}rr>{\em}r}
		\hline
		& \multicolumn{4}{c}{Excess deaths} &  \multicolumn{4}{c}{YLL} \\ \hline
		Age &  \multicolumn{2}{c}{ Dead (Other) } &  \multicolumn{2}{c}{ Dead (BC)}  &  \multicolumn{2}{c}{ Dead (Other) } &  \multicolumn{2}{c}{ Dead (BC)}  \\ 
		\hline
		&  \multicolumn{2}{c}{State 4} &  \multicolumn{2}{c}{State 5}  &  \multicolumn{2}{c}{State 4} &  \multicolumn{2}{c}{State 5}  \\ 
		\hline
		& M & S-M & M & S-M & M & S-M & M & S-M \\ 
		\hline
				S1 &  & &  &  & & &  &  \\ 
 65--69 & 363 & 362 &   0 &  -1 & 7003 & 6990 &  -7 & -19 \\ 
 70--74 & 608 & 608 &  $\mathit{-1}$ &  $\mathit{-1}$ & 9301 & 9308 &  $\mathit{-9 }$& $\mathit{-15}$ \\ 
 75--79 & 1012 & 1012 &  $\mathit{-1}$ &  $\mathit{-1}$ & 11768 & 11770 & $\mathit{-14 }$& $\mathit{-12}$ \\ 
 80--84 & 1700 & 1700 & $ \mathit{-3}$ &  $\mathit{-3} $& 14351 & 14348 & $\mathit{-22 }$& $\mathit{-25 }$\\ 
 85--89 & 2255 & 2255 & $ \mathit{-4}$ & $ \mathit{-5}$ & 13168 & 13169 & $\mathit{-24} $& $\mathit{-29} $\\ 
		S2 &  & &  &  & & &  &  \\ 
65--69 & 363 & 362 &   7 &   8 & 7001 & 6990 & 130 & 154 \\ 
70--74 & 607 & 608 &   6 &   8 & 9298 & 9308 &  99 & 122 \\ 
 75--79 & 1011 & 1012 &   7 &   9 & 11763 & 11770 &  81 & 105 \\ 
 80--84 & 1699 & 1699 &   7 &   9 & 14343 & 14340 &  56 &  76 \\ 
  85--89 & 2253 & 2254 &   4 &   6 & 13159 & 13163 &  26 &  35 \\ 
		\hline
	\end{tabular}
\end{table}

\begin{table}[H]
	\centering
	\caption{
		{ Age-specific excess number of deaths and years of life expectancy lost (YLL), per 100,000 women, based on Markov (M) and semi-Markov (S-M) models in the pandemic scenarios, Scenario 1 (S1) and Scenario 2 (S2), as compared to the pre-pandemic calibration, for {$\alpha=0.6$, $\mu^{13} = \frac{1}{7}\mu^{23}$, $\mu^{35}= 1.2 \mu^{35}$}. 
			\label{tab:DifferencesbyAge_Case3b}} 
	}
	\begin{tabular}{lr>{\em}rr>{\em}rr>{\em}rr>{\em}r}
		\hline
		& \multicolumn{4}{c}{Excess deaths} &  \multicolumn{4}{c}{YLL} \\ \hline
		Age &  \multicolumn{2}{c}{ Dead (Other) } &  \multicolumn{2}{c}{ Dead (BC)}  &  \multicolumn{2}{c}{ Dead (Other) } &  \multicolumn{2}{c}{ Dead (BC)}  \\ 
		\hline
		&  \multicolumn{2}{c}{State 4} &  \multicolumn{2}{c}{State 5}  &  \multicolumn{2}{c}{State 4} &  \multicolumn{2}{c}{State 5}  \\ 
		\hline
		& M & S-M & M & S-M & M & S-M & M & S-M \\ 
		\hline
		
S1 &  & &  &  & & &  &  \\ 
65--69 & 363 & 363 &   0 &   0 & 7003 & 7010 &  -9 &   0 \\ 
70--74 & 608 & 608 &  $\mathit{-1} $&  $\mathit{-1 }$ & 9301 & 9308 & $\mathit{-12} $& $\mathit{-15 }$\\ 
75--79 & 1012 & 1012 &  $\mathit{-2}$ &  $\mathit{-2} $& 11767 & 11770 & $\mathit{-18 }$& $\mathit{-23} $\\ 
80--84 & 1700 & 1700 &  $\mathit{-3} $& $ \mathit{-4 }$& 14350 & 14348 & $\mathit{-28} $& $\mathit{-34 }$\\ 
85--89 & 2255 & 2255 & $ \mathit{-5}$ & $ \mathit{-7}$ & 13167 & 13169 & $\mathit{-29} $&$ \mathit{-41}$ \\ 
S2 &  & &  &  & & &  &  \\ 
65--69 & 362 & 362 &   9 &  12 & 7000 & 6990 & 170 & 232 \\ 
70--74 & 607 & 608 &   8 &  10 & 9297 & 9308 & 125 & 153 \\ 
75--79 & 1011 & 1012 &   9 &  11 & 11761 & 11770 & 101 & 128 \\ 
80--84 & 1699 & 1699 &   8 &  11 & 14340 & 14340 &  68 &  93 \\ 
85--89 & 2253 & 2253 &   5 &   7 & 13156 & 13158 &  31 &  41 \\ 
		\hline
	\end{tabular}
\end{table}

\begin{landscape}
	
\section{Cancer survival at different age groups in Section~\ref{sec:Sensitivitykappa}}  \label{app:CancerSurvival_Case3}

	\begin{table}[H]
		\centering
		\caption{{1-, 5-, and 10-year survival probabilities (\%) for women from breast cancer.}}
		\label{tab:CancerSpecificSurvRatesbyAge_Case3a} 	
		\stackunder{
				\begin{tabular}{ lr>{\em}rr>{\em}rr>{\em}rr>{\em}rr>{\em}rr>{\em}rrrr}
					\hline
					&      \multicolumn{15}{c}{{Cancer Survival}}   \\
					\hline
					&          \multicolumn{6}{c}{{From State 1}}  &          \multicolumn{6}{c}{{From State 2}}   &    \multicolumn{3}{c}{{From State 3}}  \\
					\hline
					Age &  \multicolumn{2}{c}{{1-year}} &  \multicolumn{2}{c}{{5-year}} &  \multicolumn{2}{c}{{10-year}}  & \multicolumn{2}{c}{{1-year}} &  \multicolumn{2}{c}{{5-year}} &  \multicolumn{2}{c}{{10-year}} & 1-year& 5-year& 10-year\\ 
					\hline
					&   M & S-M & M & S-M & M & S-M & M & S-M & M & S-M & M & S-M & M & M & M   \\
					\hline
					&   \multicolumn{15}{c}{{ONS approach}}  \\
65--69 & 99.80 & 99.87 & 96.18 & 94.63 & 88.74 & 85.84 & 98.63 & 99.11 & 78.12 & 71.93 & 47.69 & 39.94 & 79.82 & 31.98 & 10.05 \\ 
70--74 & 99.74 & 99.83 & 95.45 & 93.59 & 87.14 & 83.91 & 98.28 & 98.87 & 74.10 & 66.73 & 41.71 & 33.63 & 74.83 & 22.85 & 5.06 \\ 
75--79 & 99.72 & 99.82 & 95.04 & 93.01 & 86.03 & 82.56 & 98.10 & 98.75 & 71.99 & 64.06 & 38.38 & 30.31 & 72.33 & 18.85 & 3.37 \\ 
80--84 & 99.65 & 99.77 & 94.14 & 91.75 & 83.52 & 79.54 & 97.68 & 98.47 & 67.44 & 58.39 & 32.03 & 24.26 & 66.58 & 11.79 & 1.26 \\ 
85--89 & 99.65 & 99.77 & 93.59 & 91.05 & 80.33 & 75.62 & 97.63 & 98.44 & 65.15 & 56.01 & 26.96 & 19.88 & 65.87 & 10.15 & 0.86 \\ 
					&   \multicolumn{15}{c}{{Adjusted model}}  \\
65--69 & 99.80 & 99.87 & 96.24 & 94.71 & 89.08 & 86.27 & 98.64 & 99.11 & 78.41 & 72.25 & 48.66 & 40.92 & 79.89 & 32.55 & 10.60 \\ 
70--74 & 99.75 & 99.83 & 95.57 & 93.74 & 87.83 & 84.78 & 98.29 & 98.87 & 74.68 & 67.36 & 43.42 & 35.30 & 74.97 & 23.69 & 5.61 \\ 
75--79 & 99.72 & 99.82 & 95.28 & 93.32 & 87.33 & 84.20 & 98.12 & 98.76 & 73.07 & 65.23 & 41.44 & 33.22 & 72.61 & 20.19 & 4.08 \\ 
80--84 & 99.66 & 99.78 & 94.67 & 92.43 & 86.40 & 83.14 & 97.72 & 98.49 & 69.74 & 60.83 & 37.84 & 29.56 & 67.19 & 13.69 & 1.87 \\ 
85--89 & 99.66 & 99.78 & 94.66 & 92.40 & 86.38 & 83.11 & 97.70 & 98.48 & 69.66 & 60.71 & 37.75 & 29.48 & 67.03 & 13.53 & 1.83 \\ 
					\hline
				\end{tabular}
		}
		{\parbox{8in}{
				\footnotesize Note: Results are for women with pre-metastatic and metastatic breast cancer, and for women with undiagnosed breast cancer based on Markov (M) and semi-Markov (S-M) models, {using  \eqref{eq:CancerSurvival}}, in the pre-pandemic calibration for $\alpha=0.6$, $\mu^{13} = \frac{1}{7}\mu^{23}$, $\mu^{35}= 0.8 \mu^{35}$.
			}
		}
	\end{table}

	\begin{table}[H]
	\centering
	\caption{{1-, 5-, and 10-year survival probabilities (\%) for women from breast cancer.}}
	\label{tab:CancerSpecificSurvRatesbyAge_Case3b} 	
	\stackunder{
			\begin{tabular}{ lr>{\em}rr>{\em}rr>{\em}rr>{\em}rr>{\em}rr>{\em}rrrr}
				\hline
				&      \multicolumn{15}{c}{{Cancer Survival}}   \\
				\hline
				&          \multicolumn{6}{c}{{From State 1}}  &          \multicolumn{6}{c}{{From State 2}}   &    \multicolumn{3}{c}{{From State 3}}  \\
				\hline
				Age &  \multicolumn{2}{c}{{1-year}} &  \multicolumn{2}{c}{{5-year}} &  \multicolumn{2}{c}{{10-year}}  & \multicolumn{2}{c}{{1-year}} &  \multicolumn{2}{c}{{5-year}} &  \multicolumn{2}{c}{{10-year}} & 1-year& 5-year& 10-year\\ 
				\hline
				&   M & S-M & M & S-M & M & S-M & M & S-M & M & S-M & M & S-M & M & M & M   \\
				\hline
				&   \multicolumn{15}{c}{{ONS approach}}  \\
65--69 & 99.71 & 99.81 & 95.06 & 93.01 & 86.69 & 83.41 & 98.02 & 98.70 & 71.93 & 63.81 & 39.54 & 31.34 & 71.32 & 18.15 & 3.24 \\ 
 70--74 & 99.63 & 99.76 & 94.28 & 91.89 & 85.28 & 81.76 & 97.53 & 98.37 & 67.82 & 58.45 & 34.93 & 26.81 & 64.73 & 11.02 & 1.18 \\ 
 75--79 & 99.60 & 99.73 & 93.86 & 91.28 & 84.19 & 80.45 & 97.29 & 98.20 & 65.69 & 55.81 & 32.26 & 24.34 & 61.53 & 8.34 & 0.66 \\ 
 80--84 & 99.51 & 99.68 & 92.93 & 89.98 & 81.75 & 77.55 & 96.72 & 97.81 & 61.34 & 50.49 & 27.26 & 19.94 & 54.37 & 4.24 & 0.17 \\ 
 85--89 & 99.50 & 99.67 & 92.25 & 89.12 & 78.12 & 73.15 & 96.65 & 97.77 & 58.78 & 47.95 & 22.48 & 15.97 & 53.55 & 3.55 & 0.11 \\ 				
				&   \multicolumn{15}{c}{{Adjusted model}}  \\
65--69 & 99.71 & 99.81 & 95.14 & 93.11 & 87.11 & 83.94 & 98.03 & 98.70 & 72.28 & 64.19 & 40.53 & 32.28 & 71.41 & 18.57 & 3.45 \\ 
70--74 & 99.64 & 99.76 & 94.44 & 92.09 & 86.08 & 82.78 & 97.55 & 98.37 & 68.50 & 59.18 & 36.65 & 28.40 & 64.92 & 11.53 & 1.33 \\ 
75--79 & 99.60 & 99.74 & 94.16 & 91.67 & 85.71 & 82.36 & 97.31 & 98.21 & 66.96 & 57.13 & 35.29 & 27.09 & 61.88 & 9.07 & 0.82 \\ 
 80--84 & 99.52 & 99.68 & 93.59 & 90.83 & 85.03 & 81.62 & 96.78 & 97.84 & 63.97 & 53.15 & 32.97 & 24.95 & 55.07 & 5.07 & 0.26 \\ 
 85--89 & 99.52 & 99.68 & 93.58 & 90.81 & 85.01 & 81.60 & 96.76 & 97.83 & 63.90 & 53.05 & 32.91 & 24.91 & 54.88 & 4.98 & 0.25 \\				
				\hline
			\end{tabular}
	}
	{\parbox{8in}{
			\footnotesize Note: Results are for women with pre-metastatic and metastatic breast cancer, and for women with undiagnosed breast cancer based on Markov (M) and semi-Markov (S-M) models, {using  \eqref{eq:CancerSurvival}}, in the pre-pandemic calibration for $\alpha=0.6$, $\mu^{13} = \frac{1}{7}\mu^{23}$, $\mu^{35}= 1.2 \mu^{35}$.
		}
	}
\end{table}

\end{landscape}

\end{appendices}
\end{document}